\documentclass[aps,prl,twocolumn,superscriptaddress]{revtex4}
\usepackage{amsmath,graphicx,textcomp}
\usepackage{placeins}
\usepackage{bbm}
\usepackage{xcolor}

\begin{document}
\title{4$\pi$ periodic Andreev bound states in a Dirac semimetal}

\author{Chuan Li}
\affiliation{MESA$^+$ Institute for Nanotechnology, University of Twente, The Netherlands}
\author{Jorrit C. de Boer}
\affiliation{MESA$^+$ Institute for Nanotechnology, University of Twente, The Netherlands}
\author{Bob de Ronde}
\affiliation{MESA$^+$ Institute for Nanotechnology, University of Twente, The Netherlands}
\author{Shyama V. Ramankutty}
\affiliation{Van der Waals - Zeeman Institute, IoP, University of Amsterdam, The Netherlands}
\author{Erik van Heumen}
\affiliation{Van der Waals - Zeeman Institute, IoP, University of Amsterdam, The Netherlands}
\author{Yingkai Huang}
\affiliation{Van der Waals - Zeeman Institute, IoP, University of Amsterdam, The Netherlands}
\author{Anne de Visser}
\affiliation{Van der Waals - Zeeman Institute, IoP, University of Amsterdam, The Netherlands}
\author{Alexander A. Golubov}
\affiliation{MESA$^+$ Institute for Nanotechnology, University of Twente, The Netherlands}
\author{Mark S. Golden}
\affiliation{Van der Waals - Zeeman Institute, IoP, University of Amsterdam, The Netherlands}	
\author{Alexander Brinkman}
\affiliation{MESA$^+$ Institute for Nanotechnology, University of Twente, The Netherlands}
	
\today
\begin{abstract}
Electrons in a Dirac semimetals possess linear dispersion in all three spatial dimensions, and form  part of a developing platform of novel quantum materials. Bi$_{1-x}$Sb$_x$ supports a three-dimensional Dirac cone at the Sb-induced band inversion point. Nanoscale phase-sensitive junction technology is used to induce superconductivity in this Dirac semimetal. Radio frequency irradiation experiments reveal a significant contribution of 4$\pi$-periodic Andreev bound states to the supercurrent in Nb-Bi$_{0.97}$Sb$_{0.03}$-Nb Josephson junctions. 
The conditions for a substantial $4\pi$ contribution to the supercurrent are favourable because of the Dirac cone's topological protection against backscattering, providing very broad transmission resonances. The large g-factor of the Zeeman effect from a magnetic field applied in the plane of the junction, allows tuning of the Josephson junctions from 0 to $\pi$ regimes. 
\end{abstract}

\maketitle
	
The concept of the band structure topology of solids has recently been extended from topological insulators to metallic systems. Whereas topological insulators are characterized by conducting surface or edge states and an insulating bulk possessing a semiconducting band gap, there is no band gap in the topological Dirac and Weyl semimetals. Rather, the bulk band structure shows a linear dispersion in all three $k$-directions, with a locking between the electron momentum and its spin (or orbit). By breaking time-reversal or inversion symmetry, the degenerate Dirac cone of such a semimetal can be split in reciprocal space to yield non-degenerate Weyl cones with opposite chiralities. Examples of Dirac semimetals (DSM) include Na$_3$Bi \cite{ZKLiu}, Cd$_3$As$_2$ \cite{Neupane,Borisenko}, and Bi$_{1-x}$Sb$_x$ \cite{KimPRL}. 
	
A topological material can be combined with a superconductor so as to give topological superconductivity \cite{SCZhang}. The interest in topological superconductors is largely given by the wish to combine the inherent electron-hole symmetry of the excitations in a superconductor with the helical nature of the electronic states in topological materials so as to form Majorana zero-energy states \cite{Fu2008}. The expected non-abelian statistics displayed by these zero modes should provide a way of performing topological quantum computation by braiding \cite{Sarma}. Platforms in which (signatures of) Majorana modes have been observed are semiconductors with Rashba spin-orbit interaction \cite{Kouwenhoven,Rohkinson}, ferromagnetic atom chains \cite{Stevan}, and topological insulators \cite{Sun,Wiedenmann}, all in combination with superconductors.
	
First signatures of superconductivity in Dirac semimetals have been reported, e.g. by applying pressure \cite{LPHe} or by using point contacts \cite{Aggarwal,HWang}, but topological aspects of Dirac semimetal superconductivity have not been studied. Here, we report on the realization of proximity induced superconductivity into a Dirac semimetal, and we reveal a significant contribution of 4$\pi$-periodic Andreev bound states to the supercurrent, made possible by the electron spin-momentum locking in the Dirac cone. 
	
\begin{figure*}
\includegraphics[clip=true,width=11cm]{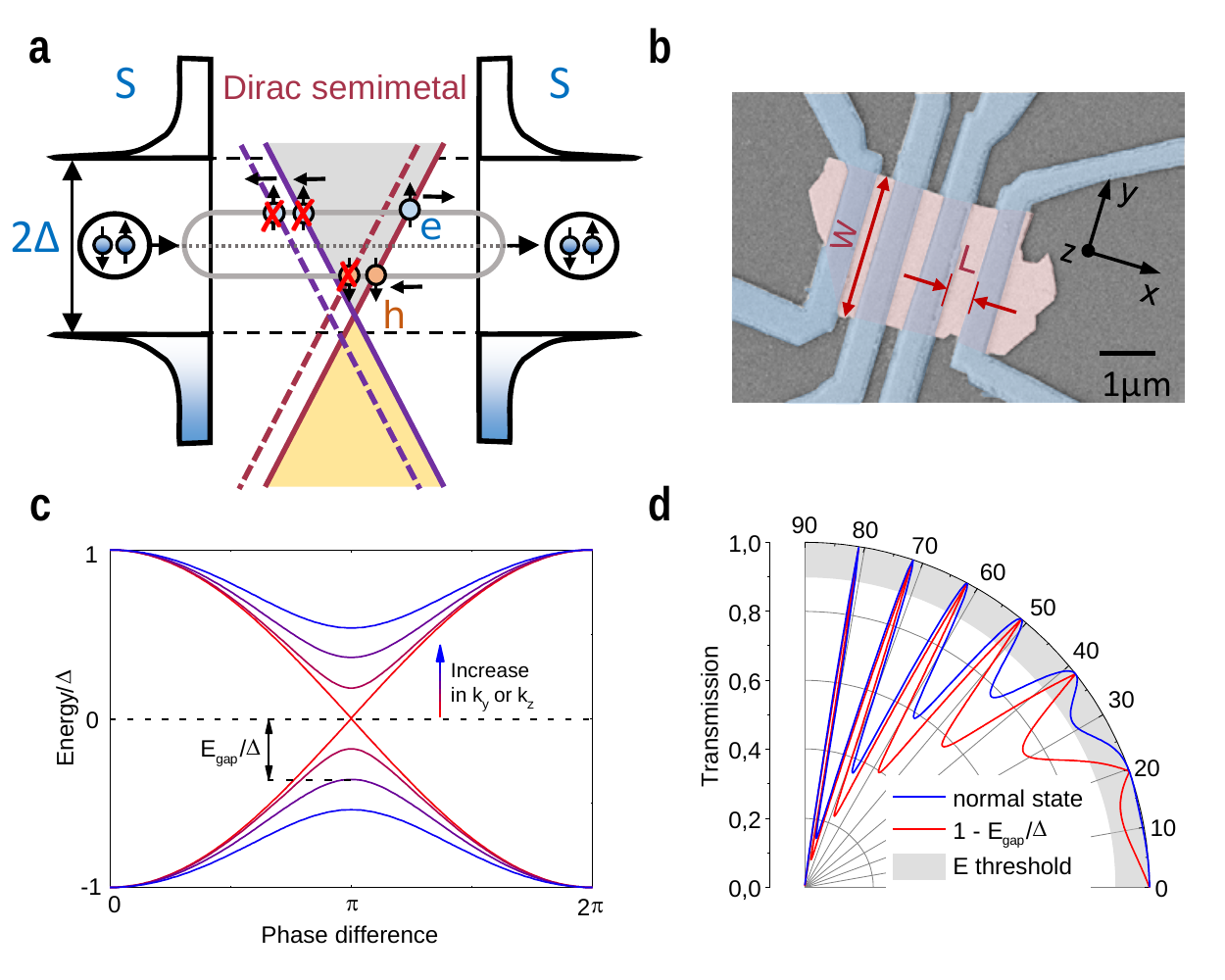}
\caption{\textbf{Josephson effect in a topological Dirac semimetal junction.} \textbf{a,} Sketch of a DSM Josephson junction, where Andreev bound states carry supercurrent from one superconducting lead (S) to the other. In the DSM interlayer, the rightgoing electron (blue, moving right in one of the Dirac cones) can be Andreev reflected at the right interface as a hole (orange, moving left) in the same Dirac cone. The hole can be reflected at the left interface into an electron. When the electron is topologically protected against backscattering (crossed out electron reflection process) and when the second Dirac cone (shown shifted and dashed for clarity) is quantum mechanically orthogonal to the first (crossed out backscattering as well as Andreev reflection into this cone) the Andreev bound states can become Majorana zero modes.  \textbf{b,} Scanning electron microscopy image of a Josephson junction with superconducting Nb electrodes on top of an exfoliated flake of the DSM Bi$_{0.97}$Sb$_{0.03}$. \textbf{c,} Schematic of the Andreev bound state energy spectrum as a function of the superconducting phase difference across a topological DSM-based Josephson junction for different values of the parallel momentum. For perpendicular modes ($k_y=k_z=0$), the gap ($E_{\textrm{gap}}$) at $\varphi = \pi$ is closed. These Andreev bound states give a $4\pi$-periodic contribution to the current-phase relation. \textbf{d,} The normal state transmission (blue line) is shown as a function of the angle between the propagation direction and the normal to the interfaces. Broad transmission resonances occur at specific angles, enabling the Andreev bound states to cross zero energy, i.e. $E_{\textrm{gap}}=0$, or $1-E_{\textrm{gap}}/\Delta=1$ (red line). When $E_{\textrm{gap}}$ is less than the resolution-determined cut-off energy, the Andreev bound states cannot be experimentally distinguished from being $4\pi$-periodic (grey area).}
\label{Fig:Fig0_spec}
\end{figure*} 
	
In a Josephson junction with a conducting interlayer, the supercurrent is carried by electron-hole bound states, as sketched in Fig. \ref{Fig:Fig0_spec}. The energy of these Andreev bound states is a function of the superconducting phase difference between the electrodes. In order to have the bound state crossing zero energy (a Majorana mode), 100\% probability of Andreev reflection is required. This seemingly unattainable condition is, in fact, guaranteed by the topological protection against backscattering in the DSM interlayer material. The electron cannot scatter back from the superconductor as an electron since these opposite-moving electron states in the Dirac cone are quantum mechanically orthogonal. The Dirac cone in the DSM is degenerate, but despite this, the electron still cannot scatter between cones due to orthogonality (see Supplementary Information for a full Bogoliubov-de Gennes model). However, the protected back scattering picture described above breaks down when scattering under a finite angle is considered \cite{Fu2008}, leading to a gap ($E_{\textrm{gap}}$) opening up around $E=0$ in the bound state spectrum, as can be seen in Fig. \ref{Fig:Fig0_spec}c. The gap closes again at propagation directions for which the normal state transmission through the device shows a resonance (Fig. \ref{Fig:Fig0_spec}d) \cite{Snelder}. For details, see Supplementary Information. Note, that the Majorana zero mode can only be detected by its tell-tale 4$\pi$ periodic current-phase relation as long as the measurement is faster than the inelastic relaxation between the Andreev bound states \cite{Badiane}, i.e. the experiment takes place at RF frequencies.

Bismuth is an interesting material in the context of both Dirac semimetals and topology. It has been recognized since the sixties that interband coupling in Bi around the L point of the bulk Brillouin zone invokes theoretical treatment in the framework of a relativistic Dirac equation in three dimensions \cite{Wolff}, although a small gap at the Dirac point remains. Upon doping Bi with Sb at the 3 to 4\% level this gap closes \cite{Dresselhaus}, providing a proper 3D Dirac semimetal that can be classified \cite{Nagaosa} as stemming from an accidental band touching. Increasing the Sb doping concentration further leads to inversion of the bands at L. At about 7\% doping (when the hole pocket at T has shifted below the Fermi energy), this material was the first discovered 3D topological insulator \cite{Hsieh_Nat2008,HsiehScience}. In this report, we focus on exfoliated single crystals with 3\% Sb doping (for methods, see Supplementary Information).
	
\begin{figure*}
\includegraphics[clip=true,width=16cm]{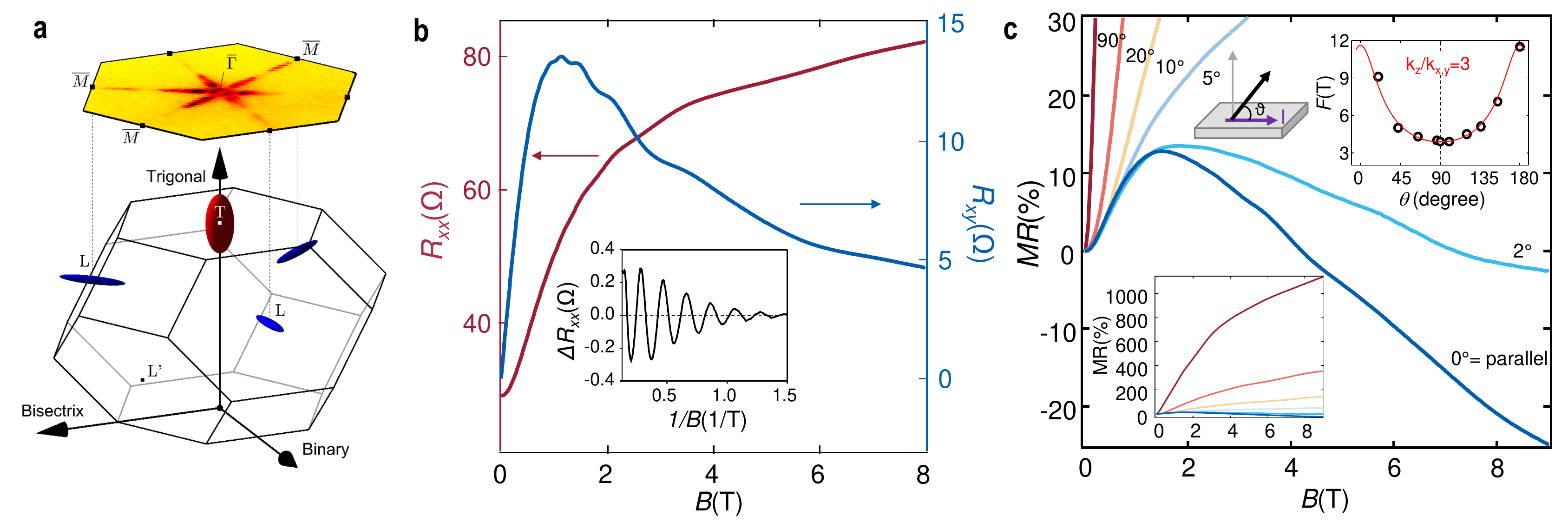}
\caption{\textbf{Magneto-transport in Bi$_{0.97}$Sb$_{0.03}$.} \textbf{a,} Schematic of the bulk Fermi surfaces of Bi$_{1-x}$Sb$_{x}$. Three electron pockets (blue) and one hole pocket (red) are located at the L and T points of the bulk Brillouin zone, respectively. The projection onto the surface Brillouin zone is also shown, including an illustrative ARPES Fermi surface map recorded from the (111) cleavage surface for $x=0.04$.  \textbf{b,} Longitudinal resistance, $R_{xx}$, and transverse Hall resistance, $R_{xy}$, as a function of perpendicular magnetic field at $T=2$ K. Inset: Shubnikov-de Haas oscillations in $R_{xx}$ after background subtraction. \textbf{c,} Angle dependence of the magnetoresistance at 2 K. The sample is rotated in the trigonal-binary plane. Negative longitudinal magnetoresistance is observed when the field is aligned with the current direction (binary axis). Upper inset: angle dependence of the Shubnikov-de Haas oscillation frequency. The fitting curve corresponds to an ellipsoidal hole pocket with elongation factor of three. Lower inset: Angle-dependence of the magnetoresistance on a zoomed-out scale. Very high magnetoresistance ($> $1000\%) is observed in perpendicular field.}
\label{Fig:Fig_MR}
\end{figure*}

The bulk Fermi surface of Bi$_{1-x}$Sb$_x$ at low doping is shown schematically in Fig. \ref{Fig:Fig_MR}a, and consists of one ellipsoidal hole pocket at the T point (whose long axis aligns with the trigonal c-axis) and three ellipsoidal electron pockets originating from the L points (whose long axes align in the trigonal-bisectrix plane, 6$^{\circ}$ tilted off the bisectrix axis). The data from angle resolved photoemission spectroscopy (ARPES) on the (111) cleavage surface of Bi$_{1-x}$Sb$_{x}$, shown in Fig. \ref{Fig:Fig_MR}a, reveal several electron and hole surface states, qualitatively very similar to what is seen at higher doping \cite{Hsieh_Nat2008,HsiehScience}. The longitudinal and transverse Hall resistance profiles in perpendicular magnetic field show Shubnikov-de Haas quantum oscillations with a period of 0.20 T$^{-1}$ (see Fig. \ref{Fig:Fig_MR}b). From the dependence on the direction of the applied magnetic field, the oscillations can be traced back to a spin-degenerate, bulk, hole ellipsoidal Fermi sheet with a measured elongation factor of three and a carrier density of $1.5\times10^{17}$ cm$^{-3}$. This density is even lower than in Bi, due to the Sb doping (see Supplementary Information). A Dingle analysis yields a hole mobility of 1.8 m$^2$V$^{-1}$s$^{-1}$ (see Supplementary Information).
		
A negative magnetoresistance is observed when the magnetic field is applied in the direction parallel to the applied electric field. This is an indication for the chiral anomaly found in Dirac semimetals \cite{KimPRL,Xiong}. We find this effect to be largest for $x=0.03$ and we measured it in exfoliated micron sized flakes with a thickness down to 100 nm, as well as in millimeter sized crystals.
	
The contribution of the Dirac cone to the conductance can be obtained from a multiband fit to the resistance data, using the ARPES (surface states) and Shubnikov-de Haas data (bulk states) as input parameters (see Supplementary Information). The surface state mobilities must be low since no quantum oscillations are observed at frequencies corresponding to their carrier densities. A bulk electron carrier density of $0.2\times10^{17}$ cm$^{-3}$ is obtained for each of the three two-fold degenerate Dirac cones. The electron mobility has a similar order of magnitude as the hole mobility and the bulk normal state conductance dominates that of the  surface because of the much higher mobility of the former. For Josephson junctions, the bulk dominance of the supercurrent will be even stronger, as the  mean free path of the electrons in the bulk Dirac cone is large enough to provide ballistic transport with a large coherence length.  
The conduction by the surface carriers and the bulk holes is diffusive, providing much shorter coherence lengths, strongly reducing these contributions to the Josephson supercurrent. (See Supplementary Information.)

Josephson junctions of varying width and Nb electrode separation length were fabricated. All junctions show a supercurrent as a clear manifestation of proximity-induced superconductivity in the Dirac semimetal samples. The junctions with the shortest length (500 nm) show ballistic Josephson transport with a critical current, $I_c$, of about 1 $\mu$A and a normal state resistance, $R_N$, of about 20 $\Omega$ (see Supplementary Information).
	
When irradiated with microwaves of frequency $\nu$, Shapiro steps are observed in the dc current-voltage curve at voltages $n\frac{h}{2e}\nu$ due to the ac Josephson effect, see Fig. 3. The shortest junctions miss the $n=1$ steps in their Shapiro spectra. This fractional Josephson effect was predicted as a detection method for the 4$\pi$-periodic current-phase relation, underlying the Majorana zero mode \cite{Fernando}. Whereas the disappearence of odd Shapiro steps have been measured occasionally \cite{Bocquillon}, experiments often just reveal the $n=1$ step to be missing \cite{Rohkinson,Wiedenmann}, which has been theoretically explained to be due to capacitive effects \cite{Dominguez}, see also the Supplementary Information.

\begin{figure*}
\includegraphics[clip=true,width=15cm]{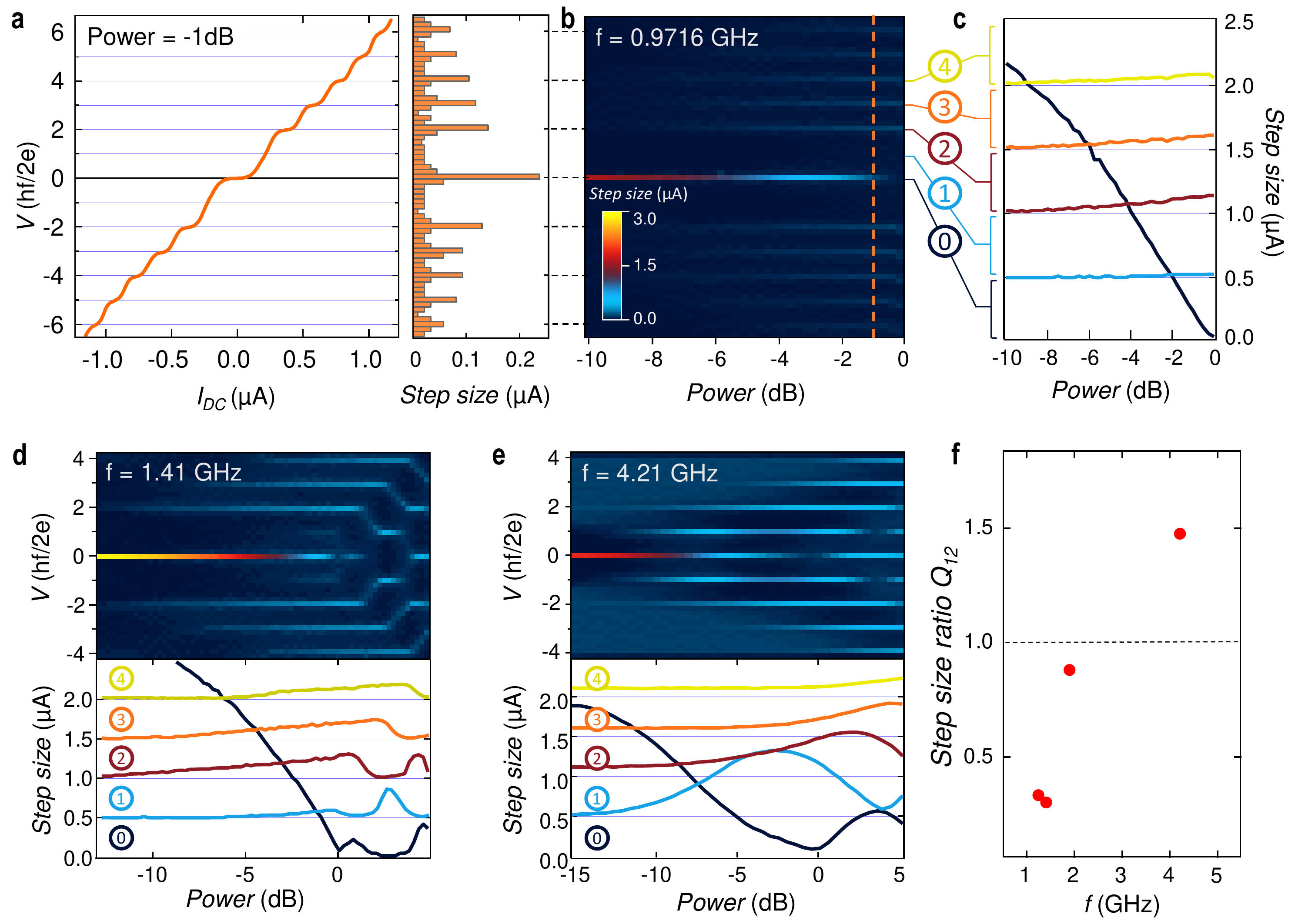}
\caption{\textbf{Fractional Josephson effect.} \textbf{a,} Current-voltage characteristics of device S1 at $T=12$ mK under microwave irradiation at frequency $\nu=0.90$ GHz and a power of -1.0 dB, where 0 dB refers to the first minimum in $I_c$. Right panel: Binning map of the Shapiro step size. \textbf{b,} Shapiro step size as function of DC voltage and RF power as derived from the binning maps. The $n=1$ step is completely suppressed in the whole range of applied microwave power. \textbf{c,} Cross-sections of panel B to reflect the power dependence of the Shapiro step sizes. For clarity, the curves are offset by 0.5 $\mu$A. The step size of $I_0$ is defined here as $2I_c$. \textbf{d,} Shapiro step size as function of RF power at frequency $\nu=1.25$ GHz. The $n=1$ step appears at high power only. \textbf{e,} Shapiro step size as function of RF power at frequency $\nu=6.40$ GHz. For this high frequency, all Shapiro steps are present. \textbf{f,} The ratio between the $n=1$ and $n=2$ step size, $Q_{12}$, as a function of RF frequency. The critical frequency where $Q_{12}=1$ is extracted to be about 2 GHz.}
\label{Fig:Fig_RF}
\end{figure*} 	

A clear missing (Fig. 3b) or reduced (Fig. 3d) $n=1$ step is observed at different RF frequencies. The missing $n=1$ step does appear when the frequency is increased (Fig. 3e). In models for resistively shunted junctions (RSJ) with simultaneous $2\pi$ and $4\pi$ components in the current-phase relation \cite{Wiedenmann}, this crossover frequency represents the point at which the contribution of the $4\pi$-periodic bound states is no longer visible, although these bound states are still present. In Fig. \ref{Fig:Fig_RF}f, we plot the ratio between the width of the first Shapiro step and the width of the second step, $Q_{12}$, as a function of the irradiation frequency and we estimate the transition frequency, $f_c$, to be about 2 GHz. We define the width of each step as the largest value it attains in their power profile. In the RSJ model, $f_c = \frac{2e}{h}R_NI_c^{4\pi}$ gives an estimate of the 4$\pi$-periodic contribution to the critical current of about $I_c^{4\pi}=0.2$ $\mu$A, which is about 20\% of the total critical current $I_c$. This large fraction can be qualitatively explained (see Supplementary Information) by the large number of Andreev bound states with a gap in the spectrum at $E=0$ that is smaller than the energy resolution of the experiment, making these effectively $4\pi$-periodic. The fact that many modes have such a small gap relates to the large width (both in energy and angle) of the normal state transmission resonances due to the topological spin-momentum locking. Because Landau-Zener tunneling due to an applied bias would be more effective at higher frequencies, while the data shows the opposite, we expect that the energy resolution in this experiment is determined by temperature, rather than bias voltage.
	
\begin{figure*}
\includegraphics[clip=true,width=14cm]{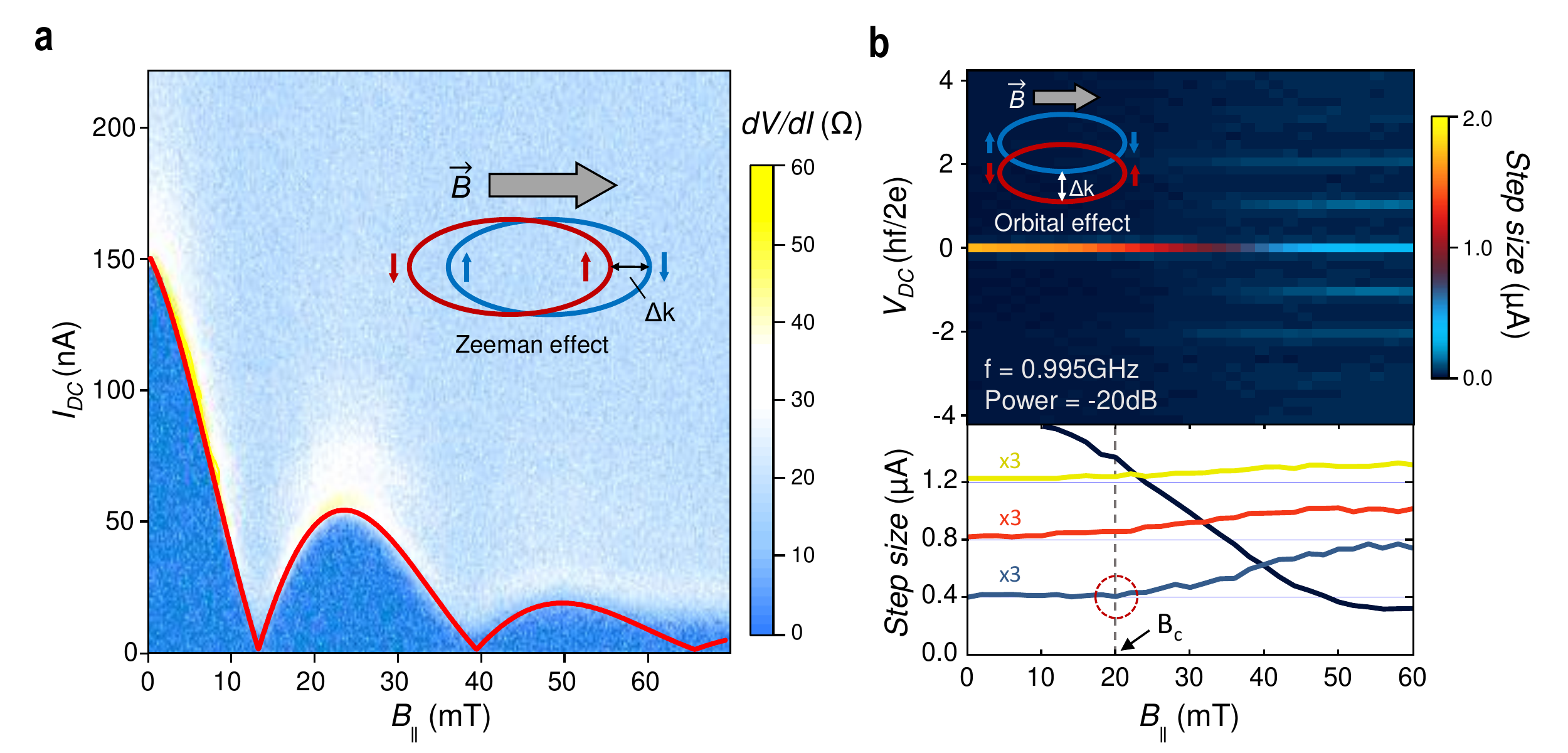}
\caption{\textbf{Finite momentum pairing.} \textbf{a,} Differential resistance of an 800 nm long junction as a function of $B$, applied parallel to the current, at $T=12$ mK. The critical current can be fitted (solid line) by considering finite momentum pairing induced by a Zeeman shift of the Dirac cone in the direction of $B$, as illustrated in the inset. \textbf{b,} Top: Shapiro step size as function of voltage and applied parallel magnetic field of a 500 nm long junction under RF irradiation ($P=-20$ dBm, $\nu=0.995$ GHz). Bottom: Cross sections of the top panel to reflect the onset of the $n=1$ Shapiro step at $B_c$, possibly due to the finite $k_y$ induced by the orbital effect of $B$ (inset top panel).}
\label{Fig:Fig4_ParallelB}
\end{figure*} 
To test our expectation that the supercurrent is carried by the electrons in the bulk Dirac cone, we study the supercurrent in a parallel magnetic field. When a field, $B_x$, is applied in the direction of the current, the Fermi surface pockets are expected to shift in $k$-space. Owing to the large $g$-factor of about 1000 for the bulk Dirac cone electrons in a magnetic field along the binary or bisectrix axes \cite{Zhu}, the Zeeman effect dominates, resulting in a shift of the Dirac cone in the $k_x$ direction of $\Delta k_x = \frac{g \mu_B B_x}{\hbar v_F}$. The proximity induced Cooper pairs then obtain a finite momentum, which is expected to lead to a spatially oscillating order parameter. Note, that at zero magnetic field, the Cooper pairs do not have a finite momentum as pairing occurs between electrons with opposite momenta at L and -L. Finite momentum pairing is known to occur in Josephson junctions with ferromagnets \cite{SFS} and, more recently, semiconductors with spin-orbit coupling \cite{Hart}. We observe an oscillating critical supercurrent as a function of the parallel magnetic field, see Fig. \ref{Fig:Fig4_ParallelB}a. In the Supplementary Information the periodicity of the oscillations is contrasted to the observed Fraunhofer pattern for perpendicular magnetic fields, ruling out contamination with perpendicular field components. Furthermore, the data can be well described using a complex coherence length, which contains the oscillations as well as the decay, as expected in the finite momentum pairing scenario \cite{Demler}. Such a simulation of the data (red line in Fig. 4a), yields critical current oscillations with a period that is set by $L \Delta k_x = \pi$, where $L$ is the length of the junction. In order to fit the data, and by using an average Fermi velocity of $5 \times 10^5$ m/s, a $g$ factor of 800 is obtained, consistent with literature \cite{Zhu}. The junction is tuned into the $\pi$-state for a parallel magnetic field between 12 and 38 mT.

In Fig. 4b we show the sensitivity of the 4$\pi$-periodic Andreev bound states to the applied magnetic field. The parallel field geometry is a convenient platform for this. For perpendicular field, the orbital contribution of the field provides a modulation of the critical current on a small field scale, which decreases the 4$\pi$-periodic Andreev bound state visibility (lowered $I_c$) before the 4$\pi$-periodicity itself is actually suppressed. However, in the parallel field orientation, we observe a reappearance of the $n=1$ Shapiro step at $B_c=20$ mT for a 500 nm junction, signalling a suppression of the 4$\pi$-periodicity, well before $I_c$ is suppressed by the finite momentum pairing (see $I_0$ of the same junction in the lower panel of Fig. 4b). We speculate that the suppression of the 4$\pi$ contribution in a parallel field is due to the shift of the Dirac cone in the $k_y$ direction, caused by the orbital contribution of the field, $\Delta k_y = \frac{e}{\hbar}B_x z$. This shift is much smaller than the Zeeman shift but it does give the Andreev bound states considerable momentum parallel to the interface, changing the angles at which transmission resonances occur. By taking $z=300$ nm as the thickness of our flake, we extract a parallel momentum of the order of $10^7$ m$^{-1}$ at $B_c=20$ mT, which indeed is significant with respect to the forward electron momentum of $3\times 10^7$ m$^{-1}$. 
	
Our observation of proximity induced superconductivity in a topological Dirac semimetal provides a platform to investigate whether topological superconductors generated in this manner have an unconventional order parameter symmetry and opens up a new avenue towards topological quantum computation. The degeneracy in the Dirac cone and the presence of multiple Dirac cones allows for multiple Majorana zero modes. It will be intriguing to see whether in the future it would be advantageous to engineer the number of cones using thin film technology \cite{DresselhausNano} or empty cones by means of valley polarization \cite{ZZhu}, and whether multiple Majoranas can be employed in quantum algorithms. Technologically, the use of the topological bulk properties of a semimetal rather than a topological surface, renders devices less sensitive to disorder and environmentally-induced surface degredation.

We thank Diamond Light Source for access to beamline I05 (proposal number 12969) that contributed to the results presented here, and Adrian Schiphorst, Timur Kim and Moritz Hoesch for assistance with the ARPES experiments. This work was financially supported by the Foundation for Fundamental Research on Matter (FOM), associated with the Netherlands Organization for Scientifc Research (NWO), and the European Research Council (ERC) through a Consolidator Grant. 


\begin{thebibliography}{1}
\bibitem{ZKLiu} Liu, Z. K. \textit{et al.} Discovery of a Three-Dimensional Topological Dirac Semimetal, Na$_3$Bi. \emph{Science} \textbf{343}, 864-867 (2014).
\bibitem{Neupane} Neupane, M. \textit{et al.} Observation of a three-dimensional topological Dirac semimetal phase in high-mobility Cd$_3$As$_2$. \emph{Nature Comm.} \textbf{5}, 3786 (2014).
\bibitem{Borisenko} Borisenko, S. \textit{et al.} Experimental Realization of a Three-Dimensional Dirac Semimetal. \emph{Phys. Rev. Lett.} \textbf{113}, 027603 (2014).
\bibitem{KimPRL} Kim, H. J. \textit{et al.} Dirac versus Weyl Fermions in Topological Insulators: Adler-Bell-Jackiw Anomaly in Transport Phenomena. \emph{Phys. Rev. Lett.} \textbf{111}, 246603 (2013).
\bibitem{SCZhang} Qi, X. L. \& Zhang, S. C. Topological insulators and superconductors. \emph{Rev. Mod. Phys.} \textbf{83}, 1057 (2011). 
\bibitem{Fu2008} Fu, L. \& Kane, C. L. Superconducting Proximity Effect and Majorana Fermions at the Surface of a Topological Insulator. \emph{Phys. Rev. Lett.} \textbf{100}, 096407 (2008).
\bibitem{Sarma} Das Sarma, S., Freedman, M. \& Nayak, C. Majorana zero modes and topological quantum computation. \emph{npj Quantum Information} \textbf{1}, 15001 (2015).
\bibitem{Kouwenhoven} Mourik, V. \textit{et al.} Signatures of Majorana Fermions in Hybrid Superconductor-Semiconductor Nanowire Devices. \emph{Science} \textbf{336}, 1003-1007 (2012).
\bibitem{Rohkinson} Rokhinson, L. P., Liu, X. \& Furdyna, J. K. The fractional a.c. Josephson effect in a semiconductor-superconductor nanowire as a signature of Majorana particles. \emph{Nature Phys.} \textbf{8}, 795-799 (2012).
\bibitem{Stevan} Nadj-Perge, S. \textit{et al.} Observation of Majorana fermions in ferromagnetic atomic chains on a superconductor. \emph{Science} \textbf{31}, 602 (2014).
\bibitem{Wiedenmann} Wiedenmann, J. \textit{et al.} 4$\pi$-periodic Josephson supercurrent in HgTe-based topological Josephson junctions. \emph{Nature Comm.} \textbf{7}, 10303 (2016).
\bibitem{Sun} Sun, H. H. \textit{et al.} Majorana Zero Mode Detected with Spin Selective Andreev Reflection in the Vortex of a Topological Superconductor. \emph{Phys. Rev. Lett.} \textbf{116}, 257003 (2016).
\bibitem{LPHe} He, L. \textit{et al.} Pressure-induced superconductivity in the three-dimensional topological Dirac semimetal Cd$_3$As$_2$. \emph{npj Quantum Mater.} \textbf{1}, 16014 (2016).
\bibitem{Aggarwal} Aggarwal, L. \textit{et al.} Unconventional superconductivity at mesoscopic point contacts on the 3D Dirac semimetal Cd$_3$As$_2$. \emph{Nature Mater.} \textbf{15}, 32-37 (2016).
\bibitem{HWang} Wang, H. \textit{et al.} Observation of superconductivity induced by a point contact on 3D Dirac semimetal Cd$_3$As$_2$ crystals. \emph{Nature Mater.} \textbf{15}, 38-42 (2016).
\bibitem{Snelder} Snelder, M. \& Veldhorst, M. \& Golubov, A. A. \& Brinkman, A. Andreev bound states and current-phase relations in three-dimensional topological insulators, \emph{Phys. Rev. B} \textbf{87}, 104507 (2013).
\bibitem{Badiane} Badiane, D. M., Houzet, M. \& Meyer, J. S. Nonequilibrium Josephson Effect through Helical Edge States. \emph{Phys. Rev. Lett.} \textbf{107}, 177002 (2011).
\bibitem{Wolff} Wolff, P. A. Matrix elements and selection rules for the two-band model of bismuth. \emph{J. Phys. Chem. Solids} \textbf{25}, 1057-1068 (1964).
\bibitem{Dresselhaus} Tichovoisky, E. J. \& Mavroides, J. G. Magnetoreflection studies on the band structure of bismuth-antimony alloys. \emph{Solid State Comm.} \textbf{7}, 927-931 (1969).
\bibitem{Nagaosa} Yang, B. J. \& Nagaosa, N. Classification of stable three-dimensional Dirac semimetals with nontrivial topology. \emph{Nature Comm.} \textbf{5}, 4898 (2014).
\bibitem{Hsieh_Nat2008} D. Hsieh \textit{et al.}, A topological Dirac insulator in a quantum spin Hall phase. \emph{Nature} \textbf{452}, 970-974 (2008).
\bibitem{HsiehScience} Hsieh, D. \textit{et al.} Observation of unconventional quantum spin textures in topological insulators. \emph{Science} \textbf{323}, 919-922 (2009).
\bibitem{Xiong} Xiong, J. \textit{et al.} Evidence for the chiral anomaly in the Dirac semimetal Na$_3$Bi. \emph{Science} \textbf{350}, 413-416 (2015).
\bibitem{Fernando} Dominguez, F., Hassler, F. \& Platero, G. Dynamical detection of Majorana fermions in current-biased nanowires. \emph{Phys. Rev. B} \textbf{86}, 140503 (2012).
\bibitem{Bocquillon} Bocquillon, E. \textit{et al.} Gapless Andreev bound states in the quantum spin Hall insulator HgTe. \emph{Nature Nanotech.} (2016), doi:10.1038/nnano.2016.159.
\bibitem{Dominguez} Pico-Cortes, J., Dominguez, F. \& Platero, G. Signatures of a 4$\pi$-periodic supercurrent in the voltage response of capacitively shunted topological Josephson junctions. \emph{cond-mat}-1703.09100 (2017).
\bibitem{Zhu} Zhu, Z., Fauqu\'{e}, B., Fuseya, Y. \& Behnia, K. Angle-resolved Landau spectrum of electrons and holes in bismuth. \emph{Phys. Rev. B} \textbf{84}, 115137 (2011).
\bibitem{SFS}  Ryazanov, V. V. \textit{et al.} Coupling of Two Superconductors through a Ferromagnet: Evidence for a $\pi$ Junction. \emph{Phys. Rev. Lett.} \textbf{86}, 2427 (2001).
\bibitem{Hart} Hart, S. \textit{et al.}, Controlled finite momentum pairing and spatially varying order parameter in proximitized HgTe quantum wells. \emph{Nature Phys.} \textbf{13}, 87-93 (2017)
\bibitem{Demler} Demler, E. A., Arnold, G. B. \& Beasley, M. R. Superconducting proximity effects in magnetic metals. \emph{Phys. Rev. B} \textbf{55}, 15174 (1997).
\bibitem{DresselhausNano} Tang, S. \& Dresselhaus, M. S. Constructing a large variety of Dirac-cone materials in the Bi$_{1-x}$Sb$_x$ thin film system. \emph{Nanoscale} \textbf{4}, 7786-7790 (2012).
\bibitem{ZZhu} Zhu, Z., Collaudin, A., Fauqu\'{e}, B., Kang, W. \& Behnia, K. Field-induced polarization of Dirac valleys in bismuth. \emph{Nature Phys.} \textbf{8}, 89-94 (2012).
\end{thebibliography}
\end{document}


\title{4$\pi$ periodic Andreev bound states in a Dirac semimetal - Supplementary information}
	
	\author{Chuan Li}
	\affiliation{MESA$^+$ Institute for Nanotechnology, University of Twente, The Netherlands}
	
	\author{Jorrit C. de Boer}
	\affiliation{MESA$^+$ Institute for Nanotechnology, University of Twente, The Netherlands}
	
	\author{Bob de Ronde}
	\affiliation{MESA$^+$ Institute for Nanotechnology, University of Twente, The Netherlands}
		
	\author{Shyama V. Ramankutty}
	\affiliation{Van der Waals - Zeeman Institute, IoP, University of Amsterdam, The Netherlands}
		
	\author{Erik van Heumen}
	\affiliation{Van der Waals - Zeeman Institute, IoP, University of Amsterdam, The Netherlands}
	
	\author{Yingkai Huang}
	\affiliation{Van der Waals - Zeeman Institute, IoP, University of Amsterdam, The Netherlands}
	
	\author{Anne de Visser}
	\affiliation{Van der Waals - Zeeman Institute, IoP, University of Amsterdam, The Netherlands}
	
	\author{Alexander A. Golubov}
\affiliation{MESA$^+$ Institute for Nanotechnology, University of Twente, The Netherlands}
	
	\author{Mark S. Golden}
	\affiliation{Van der Waals - Zeeman Institute, IoP, University of Amsterdam, The Netherlands}
	
	\author{Alexander Brinkman}
	\affiliation{MESA$^+$ Institute for Nanotechnology, University of Twente, The Netherlands}
	
	\today
	
	\begin{abstract}

A. Crystal growth and device fabrication

B. Angle Resolved Photoemission Spectroscopy

C. Shubnikov-de Haas analysis

D. Multiband fit to transport measurements

E. Josephson junctions in different regimes

F. Influence of hysteresis on the Shapiro steps 

G. Bogoliubov-de Gennes formalism for Andreev bound states in a Dirac semimetal

H. Resonances and estimated 4$\pi$ contribution 

I. Field dependence of the critical current
\end{abstract}
	
	\maketitle

\subsection{Crystal growth and device fabrication}

Bi$_{1-x}$Sb$_x$ single crystals are grown using a modified Bridgman method. High-purity Bi ingots (99.999\%) and Sb ingots (99.9999\%) were packed in a cone-shaped quartz tube and sealed under vacuum ($4 \times 10^{-7}$ mbar). The tube was first put in a box furnace and heated up to 600 $^{\circ}$C for 12 hours. The tube was shaken several times in order to obtain a homogeneous mixture of Bi and Sb. Then the tube was quickly cooled to room temperature and hung vertically in a mirror furnace for crystal growth. The tube was heated to 300-400 $^{\circ}$C, starting from the cone-shaped bottom, and the molten zone was translated up with a rate of 1 mm/hour. Flat crystals up to 1 cm in length were obtained by cleaving the crystal boule.

The Bi$_{0.97}$Sb$_{0.03}$ crystal was mechanically exfoliated onto a SiO$_2$/Si$^{++}$ substrate. The flakes are on average about 300 nm thick. The Josephson junctions are fabricated using standard e-beam lithography followed by an RF Ar$^+$ ion etch, and \textit{in situ} sputter deposition of 100 nm thick Nb electrodes and a few nm Pd as a capping layer.

\subsection{Angle Resolved Photoemission Spectroscopy}

ARPES measurements were carried out at the I05 beamline of Diamond Light Source Ltd.. Crystals of Bi$_{1-x}$Sb$_x$ ($x=0.03$ and 0.04) were cleaved in ultrahigh vacuum at a temperature of 30 K, and the ARPES data were recorded at a temperature of 10 K in the low $10^{-10}$ mbar pressure range, with an overall energy resolution of 15 meV and $k$ resolution of 0.015 \AA$^{-1}$. The Fermi surface map shown in Fig. 2 of the main paper was recorded using circularly polarised radiation of energy 60 eV from an $x=0.04$ crystal. The ARPES data from $x=0.03$ and $x=0.04$ crystals were very similar, both showing three features: an electron pocket e1 around $\bar{\Gamma}$; a hole pocket 'petal' h1 and a second electron pocket petal e2, as expected theoretically \cite{Teo}. h1 and e2 are located on a line between $\bar{\Gamma}$ and $\bar{\textrm{M}}$. ARPES studies on low Sb doping levels, have shown e1 and e2 to be from the same band \cite{Nakamura,Benia} and thus only two different surface states are present for defect-free samples. From the agreement with the of data from carefully-prepared thin films reported in Ref. \onlinecite{Benia}, it would seem our cleaves have a low defect density. 

The surface state features were the sharpest and clearest for the $x=0.04$ data, and thus these were used for the quantification of the $k_F$ values of each of the three features. Fig. S1a shows a comparison of the Fermi surface map from the $x=0.03$ (left) and $x=0.04$ (right) samples. It is evident that the Fermi surfaces of the surface states do not differ within the experimental uncertainty. Fig. S1b shows the principle $k$-space cuts used to assess the area of the 2D Fermi surfaces of the surface state features e1, h1, and e2. Table S1 contains the $k_F$ values for the long and short axes of each of the surface state charge carrier pockets, derived from the cuts through the ARPES data as shown in Fig. S1B. These are used for the simulation of the magnetoresistance data, whereby each Fermi surface contributes a 2D carrier density given by $n_{\textrm{2D}}=\frac{k_{F1}k_{F2} }{4 \pi}$ per spin direction. The degeneracy per pocket is indicated in Table S1 as well.

\begin{figure}
\includegraphics[clip=true,width=15cm]{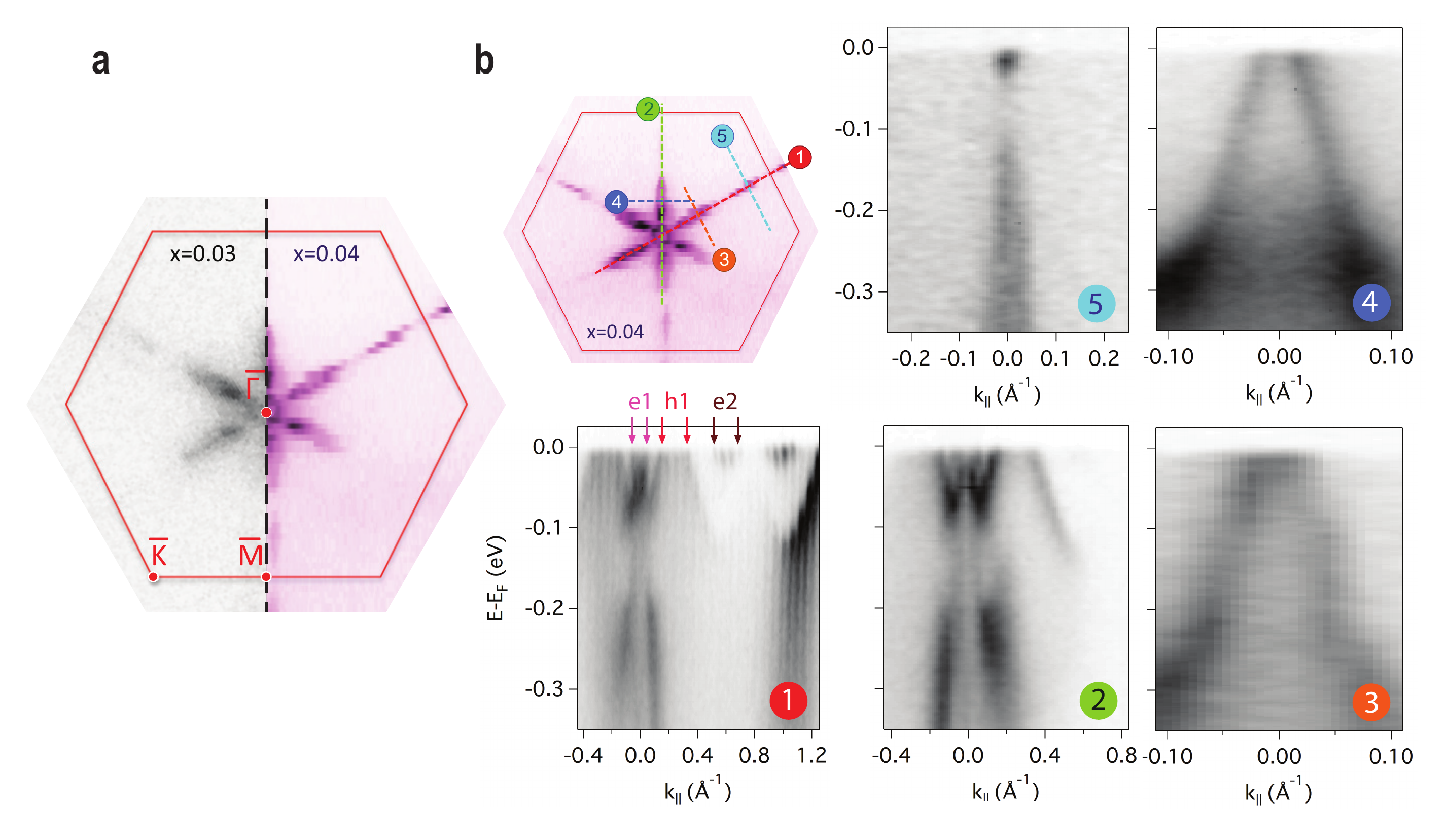}
\caption{\textbf{ARPES data from (111) cleavage surfaces of Bi$_{1-x}$Sb$_{x}$ at $x=0.03$ and $x=0.04$.} \textbf{a,} Comparison of Fermi surface maps for the two dopings, recorded with 60 eV photons at 10 K. \textbf{b,} Examples of the principle $k$-cuts from datasets on $x=0.04$ samples from which the areas of the surfaces state Fermi surfaces were determined using analyses of both energy distribution curves and momentum distribution curves. The $k_F$ indications for the $\bar{\Gamma}\bar{\textrm{M}}$ direction are sketched for cut 1, to illustrate the procedure. }
\label{Fig:FigARPES}
\end{figure} 

\begin{table}
	\caption{\label{Tab:ARPES} Estimation of the size of the surface state electron pocket (e1) at $\bar{\Gamma}$ and the hole pockets (h1) and electron pocket (e2) along $\bar{\Gamma}-\bar{\textrm{M}}$. The last column represents the degeneracy in terms of spin, number of pockets in the Brillouin zone and the two surfaces.}
	\begin{ruledtabular}
	\centering
		\begin{tabular}{|c|c|c|c|c|}
			pocket & $2k_{F1}$ (10$^8$ m$^{-1}$)  & $2k_{F2}$ (10$^8$ m$^{-1}$) & $n_{\textrm{2D}}$ (10$^{16}$ m$^{-2}$)& spin $\times$ pocket $\times$ surface\\
			\hline
			e1 &	8.8	& 8.8	& 1.5 & 1 $\times$ 1 $\times$ 2\\
			\hline
			h1 & 3.6 & 21.0 & 1.5 & $1 \times 6$ $\times$ 2\\
			\hline
			e2 & 3.8 & 22.0 & 1.7 & $1 \times 6$ $\times$ 2\\
		\end{tabular}
	\end{ruledtabular}
\end{table}

\FloatBarrier 
\subsection{Shubnikov-de Haas analysis}

\begin{figure}
\includegraphics[clip=true,width=8cm]{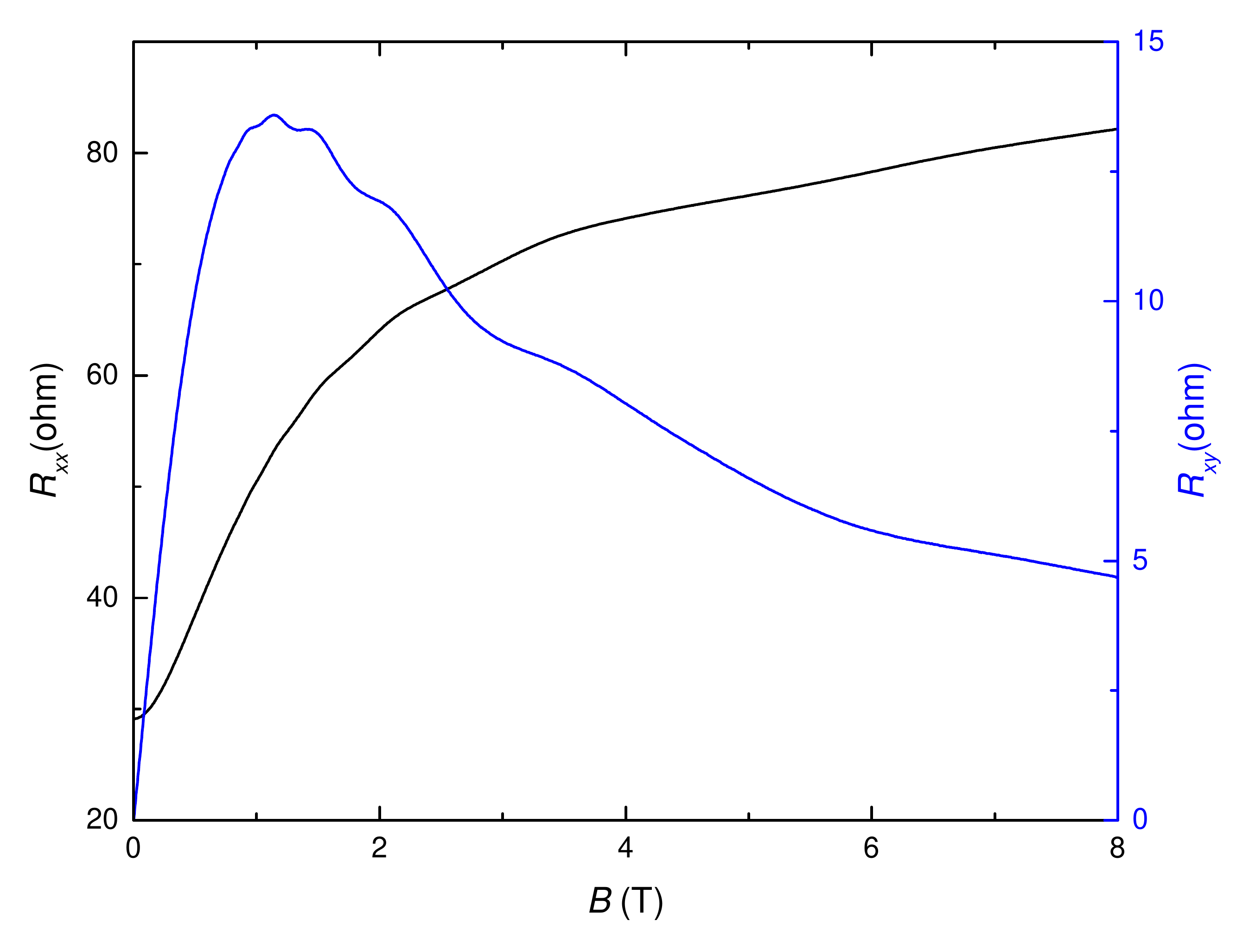}
\caption{\textbf{Typical magneto-transport data.} Symmetrized longitudinal magnetoresistance ($R_{xx}$) and Hall resistance ($R_{xy}$) of Bi$_{0.97}$Sb$_{0.03}$ flakes in perpendicular magnetic field, measured in Hall bar configuration. }
\label{Fig:FigS1}
\end{figure} 

To characterize the electronic structure of the Bi$_{0.97}$Sb$_{0.03}$ flakes, electronic transport experiments were performed and analyzed. Fig. \ref{Fig:FigS1} shows the (anti-) symmetrized data of a typical magnetoresistance measurement result. Most prominent in the figure is the nonlinearity of the Hall signal, which we observed for all our flakes. Another noticable feature is the presence of Shubnikov-de Haas oscillations, both in $R_{xx}$ and $R_{xy}$, which is very common for Bi-based materials. 

\begin{figure}
\includegraphics[clip=true,width=8cm]{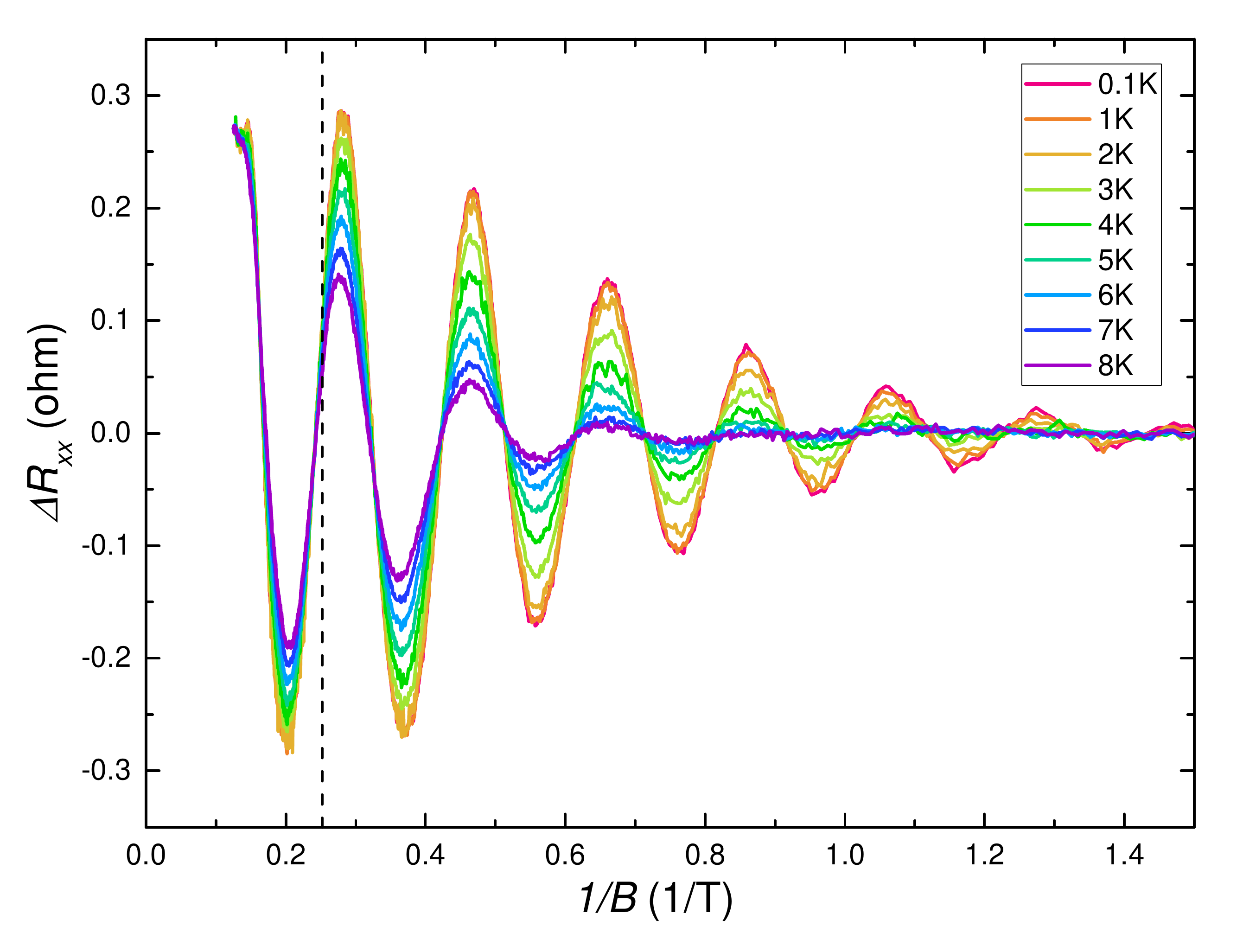}
\caption{\textbf{Temperature dependence of Shubnikov-de Haas oscillations.} Oscillations in $R_{xx}$, after background subtraction, for different temperatures. The data on the left of the dashed line are not taken into account for analysis, because of the more complicated background subtraction in this range.}
\label{Fig:FigS2}
\end{figure}

To extract the Shubnikov-de Haas oscillations most clearly from the magnetoresistance measurements, a background was subtracted from the data. Because the magnetoresistance curves have quite a complex form, we resorted to using heavily smoothed versions of the data as a background. Fig. \ref{Fig:FigS2} shows the result of such a background subtraction for different temperatures, from which we find an oscillation frequency of $F_\perp = 5.1$ T. The method we used for background subtraction becomes inaccurate for higher fields, so the amplitudes of the lowest Landau levels (those left of the dashed line in Fig. \ref{Fig:FigS2}) were excluded from further analysis.

\begin{figure}
\includegraphics[clip=true,width=8cm]{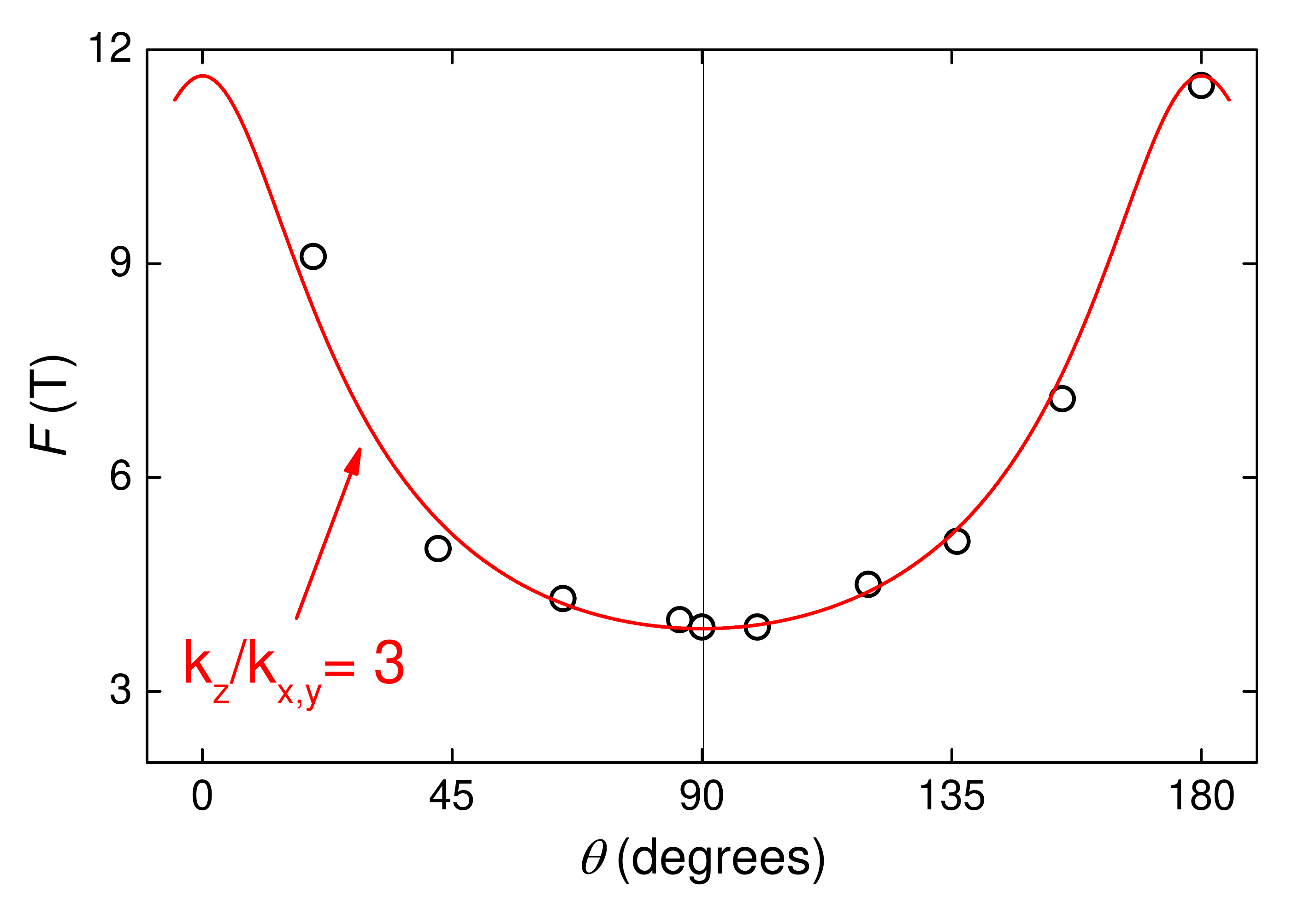}
\caption{\textbf{Angle dependence of the Shubnikov-de Haas frequency.} For different angles between the magnetic and the electric field, the Shubnikov-de Haas frequencies are plotted. The angle dependence of the frequencies fits to an ellipsoidal Fermi surface with an anisotropy factor 3.}
\label{Fig:FigS3}
\end{figure}

The period of the Shubnikov-de Haas oscillations varies with the angle between the magnetic field and the current. An example of this angle dependence is shown in Fig. \ref{Fig:FigS3}. The angle dependence of the oscillation frequency fits to that of an ellipsoidal Fermi surface, with an anisotropy factor of 3 and the long axis parallel to the c-axis of the crystal. This corresponds to the bulk hole pocket, which is located at the $T$-point (see Fig. 2a in the main paper). From $n_b^h = \frac{8 \pi}{3} 3 \sqrt{2 e F_\perp / \hbar}^3 \frac{1}{(2 \pi)^3} $, we then find a bulk hole carrier density of $n_b^h = 1.5 \times 10^{23}$ m$^{-3}$. Considering a flake thickness of 300 nm, we find a sheet carrier density of $n_b^h = 4.5 \times 10^{16}$ m$^{-2}$.

\begin{figure}
\includegraphics[clip=true,width=12cm]{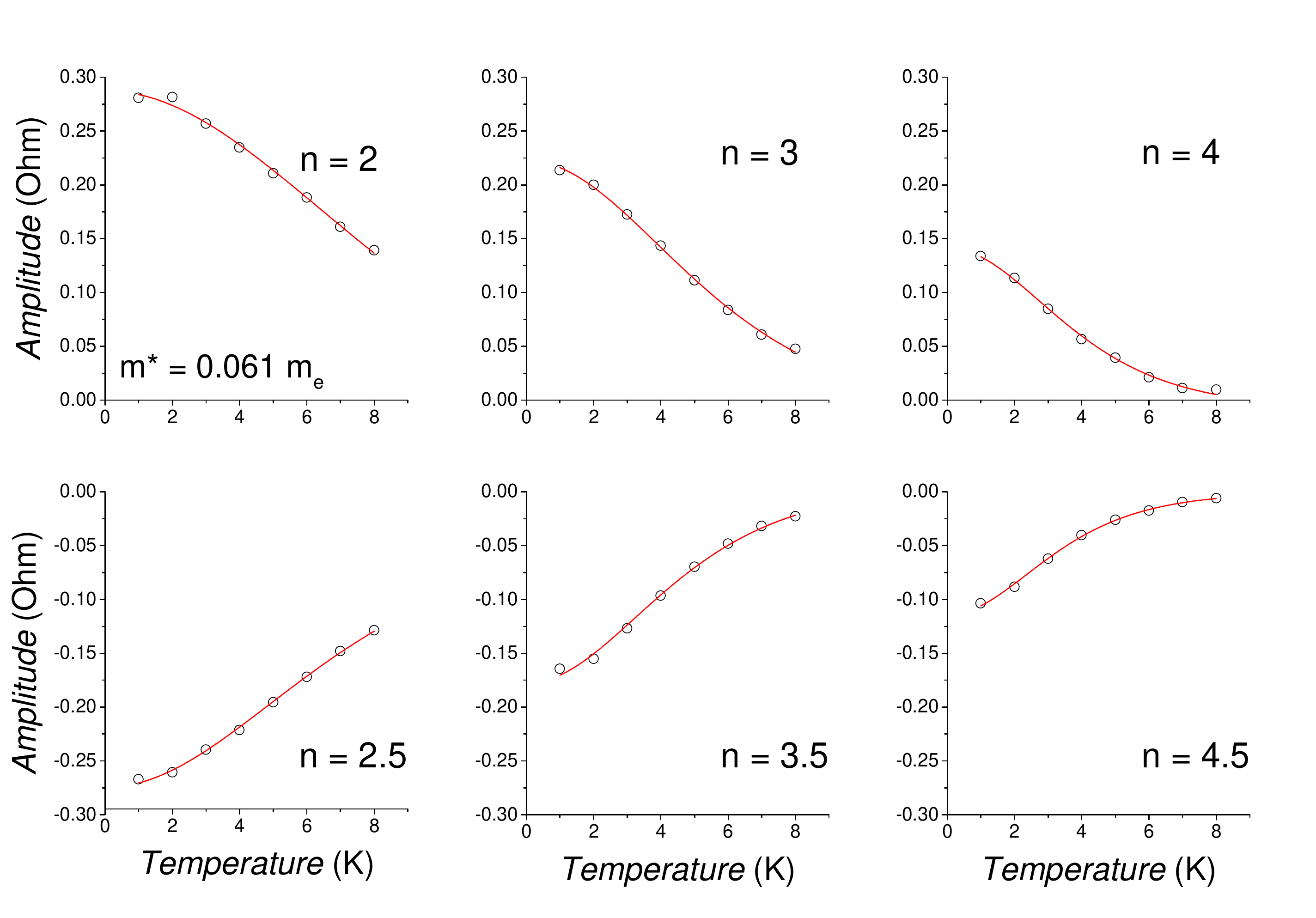}
\caption{\textbf{Lifshitz-Kosevich fits.} For six (half) integer Landau levels, the amplitudes of $\Delta R_{xx}$ as a function of temperature are fitted with the Lifshitz-Kosevich equation. From this, we find the cyclotron mass of the bulk hole carriers to be $m_{h} \approx 0.061 \, m_0$.}
\label{Fig:FigS4}
\end{figure}

\begin{figure}
\includegraphics[clip=true,width=12cm]{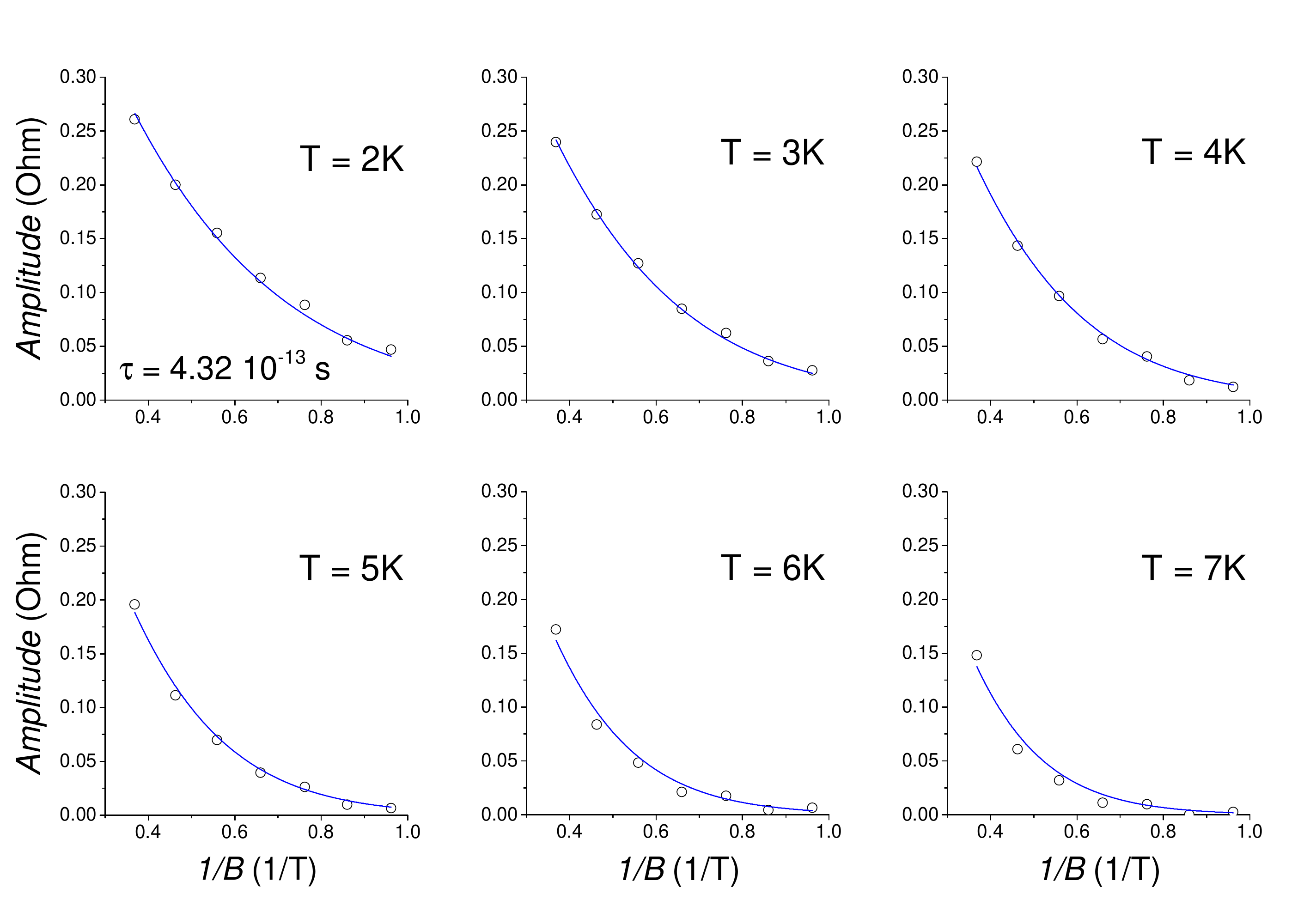}
\caption{\textbf{Dingle fits.} For six fixed temperatures, the amplitudes of $\Delta R_{xx}$ as a function of inverse magnetic field are fitted with the Dingle equation. From this, we find the quantum scattering time of the bulk hole carriers to be $\tau_q \approx 4.32  \times 10^{-13}$ s.}
\label{Fig:FigS5}
\end{figure}

The amplitude of the Shubnikov-de Haas oscillations scales with magnetic field and temperature as 
\begin{equation}
	\Delta R_{xx} \propto A_0 \frac{2 \pi^2 k_B T / E_n}{\sinh( 2 \pi^2 k_B T / E_n)} \exp{\left( -\pi \hbar / E_n \tau_q \right)},
\end{equation}
where $E_n = \hbar e B / m^{*}$. We fitted the temperature and magnetic field dependences separately by keeping the other variable fixed. Fig. \ref{Fig:FigS4} shows the dependence of the $\Delta R_{xx}$ amplitudes on temperature, for six fixed magnetic fields, corresponding to either integer or half integer Landau levels. From a global fit on all six datasets at once, we found the cyclotron mass of the bulk holes to be $m_{h} \approx 0.061 \, m_0$. To find the quantum scattering time, $\tau_q$, we used the cyclotron mass and fitted the magnetic field dependence of the amplitude for constant temperature, see Fig. \ref{Fig:FigS5}. Finally, a global fit on all six datasets at once yields $\tau_q \approx 4.32  \times 10^{-13}$ s. If we take the cyclotron mass of the bulk holes to be $m_{h} = \left(m^{*}_x m^{*}_y m^{*}_z \right)^{1/3}$ (with an anisotropy factor of 3), we find $m^{*}_x \approx 0.042 m_0$, which indicates a hole mobility of $\mu_{h} \approx 1.76$ m$^2$/V s.


\subsection{Multiband fit to transport measurements}
From the ARPES data and Shubnikov-de Haas analysis, we obtained several properties of three different types of carriers. Including the bulk electron pockets, we conclude that electronic transport in Bi$_{0.97}$Sb$_{0.03}$ is mediated by four different conduction channels. With this information, we can perform a four-band fit to the magnetoresistance data, with

\begin{equation}
	G_{xx} = \frac{e \, (n_s^e+n_s^h) \, \mu_s}{1 + (\mu_s \, B)^2} + \frac{e \, n_B^e \, \mu_B^e}{1 + (\mu_B^e \, B)^2} + \frac{e \, n_B^h \, \mu_B^h}{1 + (\mu_B^h \, B)^2}
\label{eq:Gxx}
\end{equation} 
and
\begin{equation}
	G_{xy} =  \frac{e \, (-n_s^e+n_s^h) \, \mu_s^2 \, B}{1 + (\mu_s \, B)^2} -  \frac{e \, n_B^e \, \left( \mu_B^e \right)^2 B}{1 + (\mu_B^e \, B)^2} +  \frac{e \, n_B^h \, \left( \mu_B^h \right)^2 B}{1 + (\mu_B^h \, B)^2}.
\label{eq:Gxy}
\end{equation}

\begin{figure}
\includegraphics[clip=true,width=10cm]{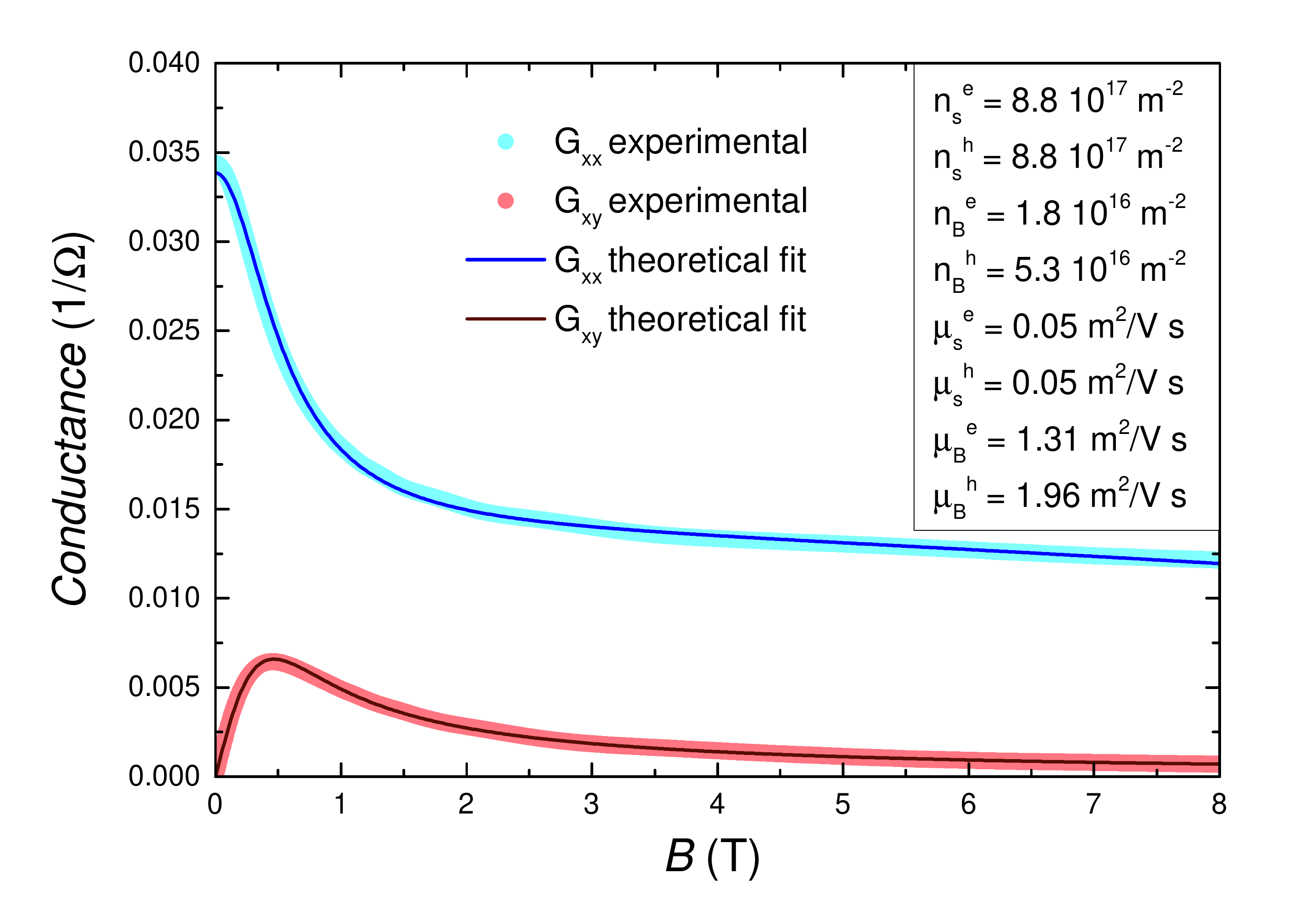}
\caption{\textbf{Multiband fit on the conductance.} Longitudinal and Hall conductance, calculated from the measured $R_{xx}$ and $R_{xy}$ and fitted with the 4-band model from Eqs. \ref{eq:Gxx} and \ref{eq:Gxy}. The parameters obtained from the fit are listed in the figure.}
\label{Fig:FigS6}
\end{figure}

\begin{figure}
\includegraphics[clip=true,width=8cm]{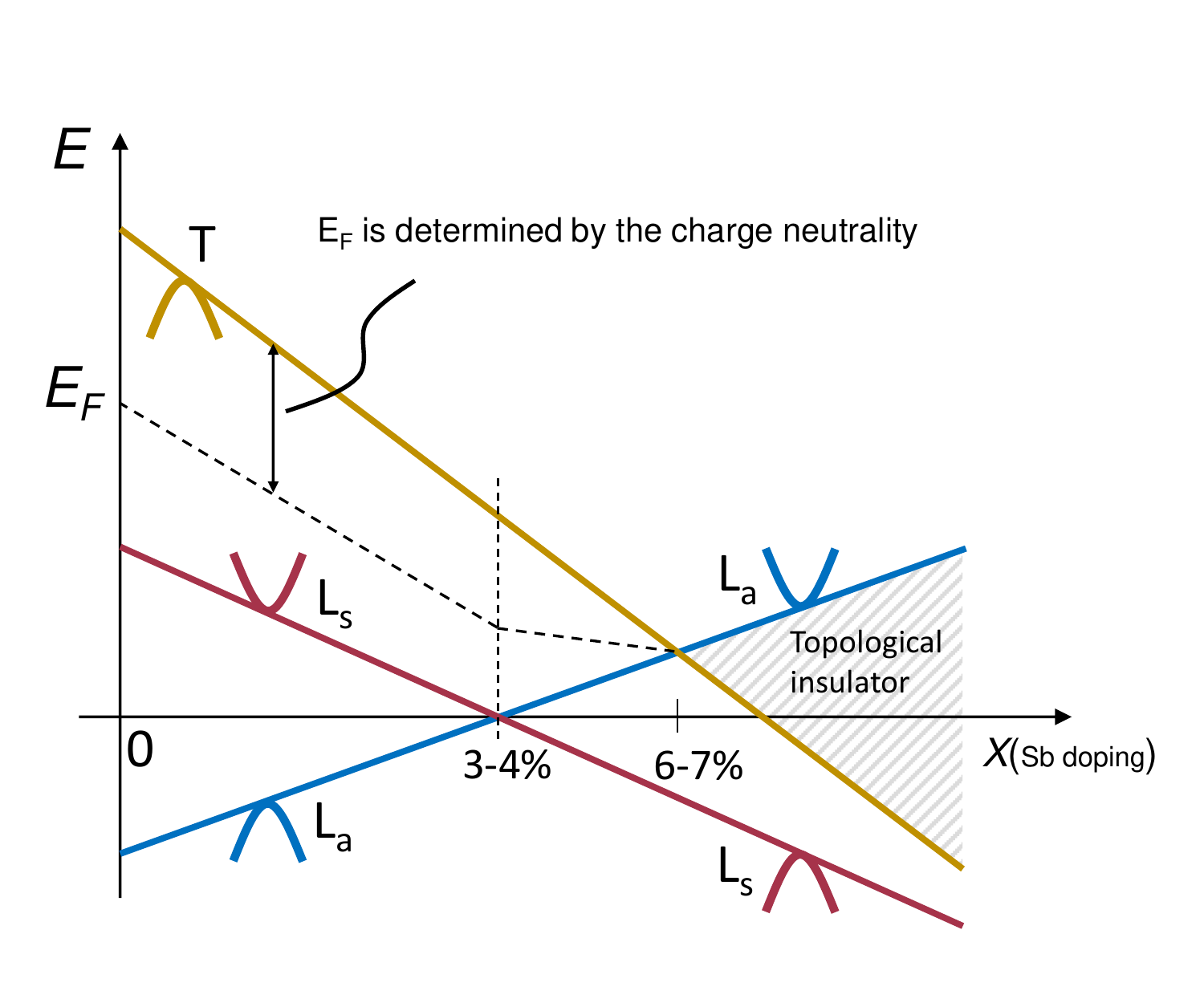}
\caption{\textbf{Sb doping dependence of the energy levels in Bi.} Schematic doping dependence of the bulk valence band at T and the valence and conduction band at L. Band inversion of L$_s$ and L$_a$ occurs around 3 to 4 \% doping of Sb. At around 6 to 7\%, the valence band at T drops below the Fermi energy, $E_F$ (dotted line), and the material becomes a topological insulator.  Figure after Ref. \onlinecite{Behnia}.}
\label{Fig:Sbdoping}
\end{figure}

The parameters that we have estimates for are $n_s^e, n_s^h, n_B^h$ and $\mu_B^h$. We also know that, since we do not see any quantum oscillations from the surface states, the surface state mobilities should be low and we assumed them equal for the electron and hole pockets. Using the estimates as fixed input parameters, we first performed an initial fit. Then we optimized the fit without any fixed parameters, although we kept all parameters within reasonable bounds. The best fits on $G_{xx} = R_{xx}/(R_{xx}^2 + R_{xy}^2)$ and $G_{xy} = R_{xy}/(R_{xx}^2 + R_{xy}^2)$ are shown in Fig. \ref{Fig:FigS6}, along with the resulting parameters. The obtained surface carrier densities are higher than the estimates from the ARPES data. We attribute this to the fact that, contrary to the crystal measured in ARPES, the measured Hall bars were exposed to air, which affected the surface states. While the surface mobilities were assumed equal, the equality of the surface carrier densities is a result from the fit. The physical consequences are charge neutrality on the surfaces and a surface contribution to the longitudinal conductance, without any significant effect on the Hall conductance. The extracted bulk electron mobility is lower than reported for pure Bi, because it is the average of the anisotropic electron mobilities. The carrier densities of the bulk carriers are lower than in pure Bi due to the Sb doping, see Fig. \ref{Fig:Sbdoping}.

From the extracted sheet electron density of the bulk of $1.8 \times 10^{16}$ m$^{-2}$ we divided by the flake thickness of 300 nm to obtain a bulk carrier density of $0.6 \times 10^{23}$ m$^{-3}$, which is only $n_e = 0.2 \times 10^{23}$ m$^{-3}$ per Dirac cone. For an ellipsoidal Fermi surface with extremal Fermi velocities $v_1=0.8 \times 10^{5}$ m/s and  $v_2=10 \times 10^{5}$ (see Ref. \onlinecite{Hsieh_Nat2008}), we obtained a Fermi energy of $E_F=\hbar \left(\pi^2 n_e v_1 v_2^2 \right)^{1/3} =16$ meV. 

\FloatBarrier

\subsection{Josephson junctions in different regimes}
The length of the Josephson junctions, as defined by the lithography process and after the Nb deposition, varies between 500 nm and 1 $\mu$m. We estimate the mobility of the Bi$_{0.97}$Sb$_{0.03}$ interlayer from the magnetotransport results of Hall bar samples made from the same crystal. The multiband and Shubnikov-de Haas analyses above yield typical $\mu$ values of order of 2 m$^2$V$^{-1}$s$^{-1}$, both for the electrons and the holes in the bulk bands. 

Bulk holes have an effective mass of $0.042m_0$ (see Shubnikov-de Haas analysis, section C) and a Fermi velocity \cite{Zhu} of $1.4 \times 10^5$ m/s. Using these values, we therefore extract an elastic mean free path, $l_e=\frac{\mu m v_F}{e}$, of about 66 nm, which is shorter than the junction length. However, for the bulk electrons with linear dispersion, the effective mass is given by \cite{Novoselov} $m^* v_F = \hbar k$, which is substantially higher along the long axis of the ellipsoidal electron Fermi surface than for the holes, giving a mean free path of the order of the length of the devices, suggesting that the shortest junctions could show ballistic transport.

\begin{figure}
\includegraphics[clip=true,width=12cm]{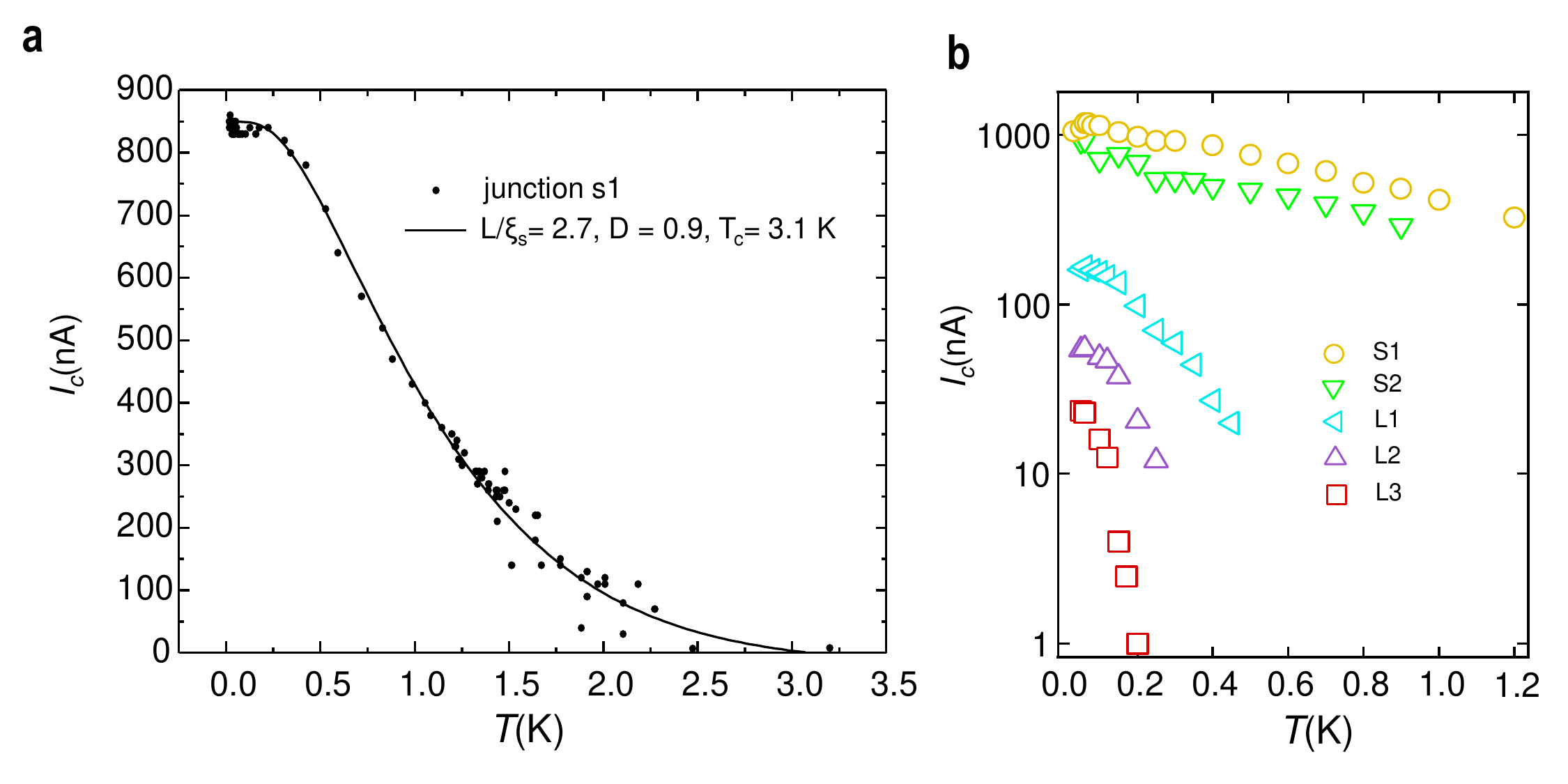}
\caption{\textbf{Critical current temperature dependence} \textbf{a,} Temperature dependence of critical current of sample S1 (data points). The temperature dependence of the critical current can be fitted by the theoretical model of ballistic Josephson junctions \cite{Galaktionov}. The parameters are shown in the figure. \textbf{b,} Temperature dependence of junctions of different length, see also Table II. While the short junctions (S1,S2) can be fitted with ballistic transport models, the longer junctions are in the diffusive limit.}
\label{Fig:Fig1_IcT}
\end{figure} 

To confirm the ballistic or diffusive behavior of the junctions, the critical current as function of temperature, $I_c(T)$, was measured for all devices. For the shortest junctions, the $I_c(T)$ function is concave, which is an indication of ballistic transport in the interlayer. We therefore fitted $I_c(T)$ with the clean limit Eilenberger equations, using the model of Galaktionov and Zaikin \cite{Galaktionov}. This model includes barriers with arbitrary transparencies between the interlayer and the superconductors. We assumed a symmetric situation with equal transparencies on both sides of the interlayer. The input parameters of the model are $T_c = 3.1$ K, as determined from the criterion that $I_c=0$ (hence $\Delta=0.5$ meV), and the device length $L=500$ nm. The model fits the data accurately for an interface transparency, $D$, of 0.9 and a normal coherence length of $\xi=\frac{\hbar v_F}{\pi \Delta}=185$ nm, giving an average $v_F=5 \times 10^5$ m/s, see Fig. \ref{Fig:Fig1_IcT}a. Note, that the $I_c(T)$ of sample S1 in Fig. \ref{Fig:Fig1_IcT}a was measured after several cooldowns, which shows a slight reduction in the critical current with respect to the first measurements, see Table S2 and Fig. \ref{Fig:Fig1_IcT}b. The overall shape, however, did not change.

The reduced junction $T_c$ with respect to the critical temperature of the Nb electrodes, as well as the high interface transparency, suggest that a proximity induced superconducting gap is induced in the Bi$_{0.97}$Sb$_{0.03}$ below the Nb electrodes. The Josephson junction is then formed laterally where the proximized regions act as superconducting electrodes. The experimentally observed deviation from a fully transparent interface can then be explained by a modest Fermi velocity mismatch (induced by the different workfunction of the Nb electrodes).

For the longer junctions, the $I_c(T)$ dependence is qualitatively different (see Fig. \ref{Fig:Fig1_IcT}b) and can only be fitted using the Usadel equations for diffusive transport \cite{Dubos}. Thus, by varying the length of the junctions, samples can be placed in different junction limit regimes. The diffusive junctions naturally have lower $R_NI_c$ values. We list all junction parameters of the different devices in Table \ref{Tab:parameters}. Only the shortest junctions, which are in the ballistic regime, show the Majorana zero modes as discussed in the main text.
\begin{table}
	\caption{\label{Tab:parameters} Parameters of all junctions }
	\begin{ruledtabular}
	\centering
		\begin{tabular}{|c|c|c|c|c|c|}
			& L1  & S1 & S2 & L2 & L3\\
			\hline
			$L$ (nm) &	800	& 500	& 500 & 800 & 1000	\\
			\hline
			$W$ ($\mu$m) & 3 & 3 & 2 & 2 & 2 \\
			\hline
			$R_N$ ($\Omega$) & 18.7 & 17.3 & 23 & 40 & 27 \\
			\hline
			$I_c$ (nA) & 160 & 1220 & 970 & 55 & 48 \\
			\hline
			$R_NI_c$ ($\mu$V) & 3 & 21.1  & 22.3 & 2.2 & 1.3 \\
		\end{tabular}
	\end{ruledtabular}
\end{table}

\FloatBarrier
\subsection{Influence of hysteresis on the Shapiro steps}

\begin{figure}
\includegraphics[clip=true,width=16cm]{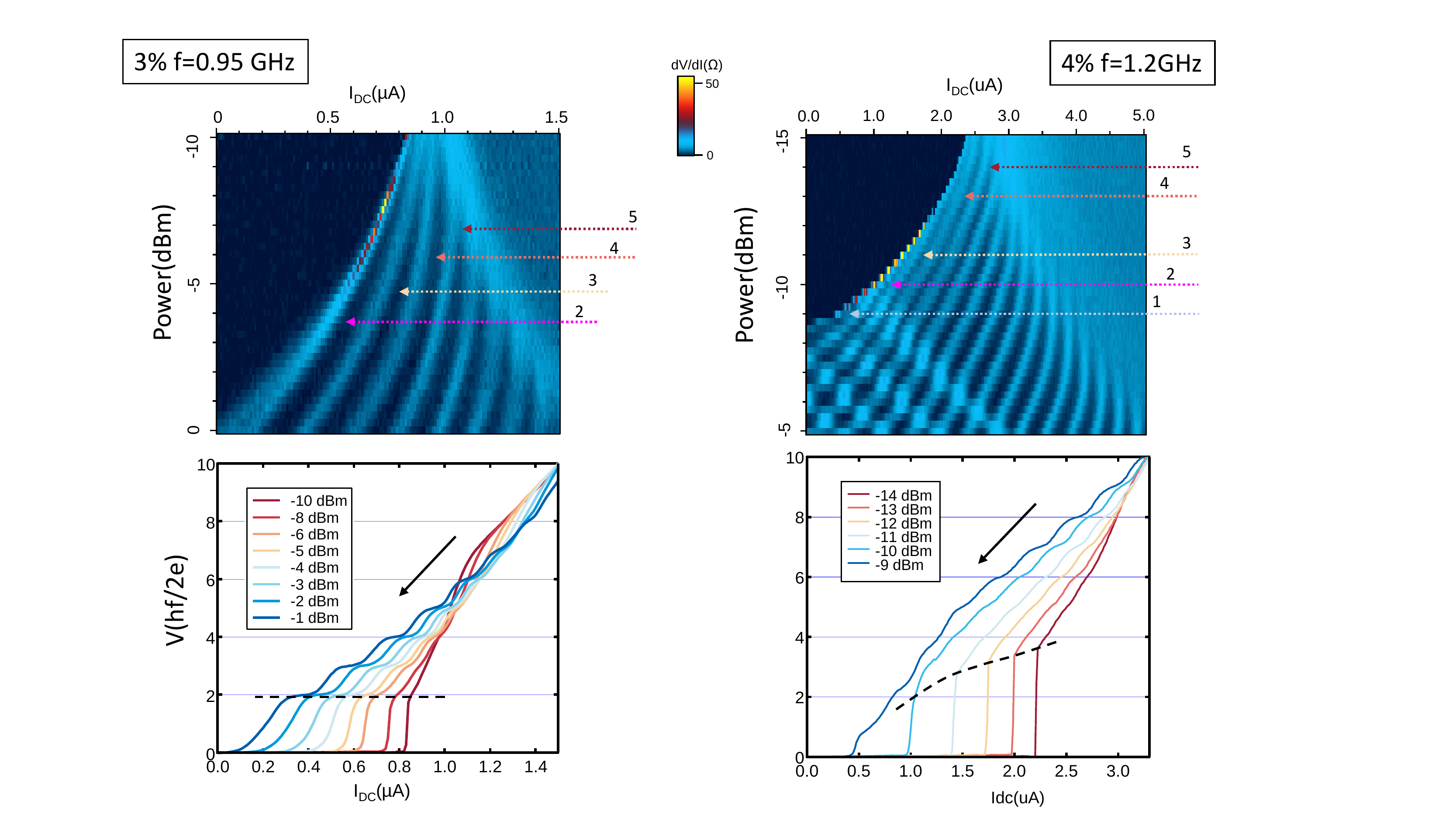}
\caption{\textbf{Comparison between 3\% and 4\% Bi$_{1-x}$Sb$_x$ junctions.} (Top) Color-plot of the differential resistance (dV/dI) in $I_{DC}$ and RF power space. The number of the steps are indicated in both color-plots. In the 4\% sample, at low RF power, all the low level steps are reduced or suppressed due to the large retrapping voltage. They all appear when the critical current is sufficiently suppressed. In the 3\% sample, the $n=1$ step is always absent even when $I_c\rightarrow 0$. (Bottom) Current-voltage characteristics (normal state $\rightarrow$ superconducting state) at different values of the RF power. The transition point is indicated by the dashed black line, which is power-dependent in the 4\% sample, but consistently positioned at the $n=2$ step for the 3\% sample.}
\label{Fig:Fig_comparison}
\end{figure}

The dynamics in a Josephson junction can be described by a resistively shunted junction (RSJ) model or by a resistively and capactivitely shunted junction (RSCJ) model, depending on the nature of the shunt. The models are characterized by a damping parameter ($\sigma$) or the so called Stewart-McCumber parameter, defined as $\beta_c=1/\sigma^2=\dfrac{2e}{\hbar}I_c R_N^2 C$. When $\beta \gtrsim 1$, the transition between the superconducting state and normal state starts to become hysteretic (retrapping current $I_r <$ switching current $I_c$) and the junction is getting into the underdamped regime. Also, if the retrapping voltage ($V_r = I_r R_N$) is larger than the jump height of a Shapiro step ($hf/2e$), some steps could be buried inside of $V_r$. It is important to distinguish our observations of 4$\pi$-bound state induced missing Shapiro steps from such effects. Here, we examine two different aspects of the RF response of the Dirac semimetal junctions to exclude the possibility of dynamic effects on our results.

\subsubsection{Comparison between 3\% and 4\% Bi$_{1-x}$Sb$_x$ junctions}
We measured both 3\% and 4\% Bi$_{1-x}$Sb$_x$-based Josephson junctions in RF regime. In Fig. \ref{Fig:Fig_comparison} we show the $dV/dI$ color plot and the $IV$ curves of two junctions with similar $I_c R_N$ product and irradiated at similar frequencies. However, the damping parameters are different, making the 4\% devices slightly hysteretic.  We indicate the number of the steps in the color-plot and it is clearly shown that for the 4\% sample, the Shapiro steps are hidden at low bias because of hysteresis and they gradually reappear when the RF power is increased. However, the Dirac semimetal (3\%) junction consistently shows a 'missing step $n=1$' through a large range of power, even when the critical current $I_c$ is reduced to nearly zero. Also, we observe that, in the high power regime, the $IV$ curves are already rounded by the heating, providing a smooth transition with finite voltage values in between the normal and superconducting states, so that the mentioned hiding effect due to the retrapping voltage can be excluded.

\subsubsection{Influence of the capacitance on the 4$\pi$-periodic supercurrent in the intermediate damping regime }
The technique of measuring missing Shapiro steps in the context of a 4$\pi$-periodic supercurrent contribution was established theoretically in main text Ref. 24, and then experimentally applied in main text Refs. 11 and 25. Our measurement results share the general trend that the lowest odd steps are suppressed more than the higher steps. In our case, only the $n=1$ step is missing. Recently, a theoretical study (main text Ref. 26) employed the RSCJ model to simulate the Shapiro steps of 2$\pi$+4$\pi$-periodic supercurrent in different regimes for junction parameters that are very much applicable to previous publications and the present work. The main finding of the theory is the effect of the capacitance in the RSCJ model on the visibility of the $4\pi$-periodic contribution to the supercurrent. First of all, the first step disappears for a large range of RF power. Secondly, the third and fifth steps are \textit{less} affected. In fact, their Fig. 2 (dashed line) shows results for a junction with $\beta_c=1/\sigma^2 \sim 1$ and a 20\% 4pi-periodic current contribution, which are the exact parameters of our junctions. Indeed, the first step is expected to be strongly suppressed while the higher odd steps are always present. We can, therefore, conclude that dynamic effects, or biasing instabilities due to hysteresis, are absent in our Dirac semimetal Josephson junctions, but that capacitive effects are still important in order to understand the degree of disappearance of different Shapiro steps.

\FloatBarrier
\subsection{Bogoliubov-de Gennes formalism for Andreev bound states in a Dirac semimetal}
A generic Hamiltonian for a three-dimensional Dirac semimetal \cite{Nagaosa} is
\begin{align}
H_D = \begin{pmatrix}
\hat{H}_0 -\hat{\mu} & 0 \\ 0 & -\hat{H}_0 -\hat{\mu}
\end{pmatrix} \label{HD},
\end{align}
where $\hat{H}_0=\hbar v \left(k_x \hat{\sigma}_x + k_y \hat{\sigma}_y+k_z \hat{\sigma}_z \right)$ and $\hat{\mu}=\mu \hat{\mathbbm{1}}$. The wave vector $\textbf{k}=\left(k_x,k_y,k_z \right)$ is measured with respect to the Dirac point (in the case of Bi$_{1-x}$Sb$_{x}$ this is the crystallographic L point in k-space), and $-\mu$ is the energy of the Dirac point with respect to the Fermi energy ($E=0$). In spherical coordinates we can write
\begin{align}
\hat{H}_0 = \hbar v k\begin{pmatrix}
\textrm{cos} \phi & e^{-i \theta} \textrm{sin} \phi  \\ e^{i \theta} \textrm{sin} \phi  & -\textrm{cos} \phi 
\end{pmatrix} \label{H0},
\end{align}
where $k=\sqrt{k_x^2+k_y^2+k_z^2}$, $\theta=\textrm{arctan} \frac{k_y}{k_x}$, and $\phi=\textrm{arccos} \frac{k_z}{k}$.

The Dirac Hamiltonian of Eq. (\ref{HD}) is a 4 $\times$ 4 Hamiltonian in a basis that is spanned by spin ($\uparrow \downarrow$) and orbital/parity (1,2) elements, i.e. $\left(u_{1\uparrow},u_{1\downarrow},u_{2\uparrow},u_{2\downarrow} \right)^T$. The Hamiltonian can be straightforwardly generalized to take anisotropy ($v$ taking different values in the three directions) or mass terms into account. In fact, in the limit of zero mass, the Hamiltonian for the bulk of a three-dimensional topological insulator such as the seminal Bi$_2$Se$_3$ \cite{Zhang} reduces to Eq. (\ref{HD}). 

The part of the band structure we consider consists of two superposed Dirac cones. When, in this case, the 4-spinor wave functions are decomposed into two independent 2-spinors, two Weyl cones are obtained with opposite Chern numbers. Here, we rather keep the 4-spinor Dirac notation since Bi$_{1-x}$Sb$_{x}$ is classified as an accidental Dirac semimetal \cite{Nagaosa} for which a gap can be opened (for example by doping away from $x=3$\%), contrary to the case of non-degenerate Weyl cones. A transition to non-degenerate Weyl cones can be made when time reversal symmetry is broken by magnetization, for example.

For energies above the Dirac point ($E>-\mu$), Eq. (\ref{HD}) provides two electron Fermi surfaces with a linear dispersion in three directions, where $E=-\mu+\hbar v k$. The spinor part of the wave functions of these two cones are orthogonal, $\psi_D^{1,2}=\frac{1}{\sqrt{2}}\left(\textrm{cos}\frac{\phi}{2},e^{i \theta}\textrm{sin}\frac{\phi}{2}, \pm \textrm{sin}\frac{\phi}{2}, \mp e^{i \theta}\textrm{cos}\frac{\phi}{2} \right)^T$, which reduces to $\psi_D^{1,2}=\frac{1}{2}\left(1,e^{i \theta}, \pm 1, \mp e^{i \theta} \right)^T$ for $k_z=0$.

We include the holes $v=u^{\dagger}$ to form a Nambu basis for the wave functions $\left(u_{1\uparrow},u_{1\downarrow},u_{2\uparrow},u_{2\downarrow},v_{1\uparrow},v_{1\downarrow},v_{2\uparrow},v_{2\downarrow}\right)^T$. In principle, exotic order parameters can be expected when inter-orbital pairing is considered \cite{Hashimoto}, but here we assume the most simple proximity induced intra-orbital s-wave singlet superconducting pairing $\Delta=\hat{\Delta}_0 \hat{\mathbbm{1}}$, where 
\begin{align}
\hat{\Delta}_0 = \begin{pmatrix}
0 & \Delta_S e^{i \varphi} \\ -\Delta_S e^{i \varphi} & 0
\end{pmatrix} \label{Delta0},
\end{align}
with $\varphi$ being the superconducting phase of the condensate. The 8 $\times$ 8 Bogoliubov-de Gennes Hamiltonian is then given by
\begin{align}
H_{BdG} = \begin{pmatrix}
H_D (\textbf{k}) & \Delta \\ -\Delta^{*} & -H_D^{*}(-\textbf{k})
\end{pmatrix} \label{BdG}.
\end{align}
The reversal of the vector $\textbf{k}$ is provided by taking $\phi \rightarrow \phi+\pi$, so that for example
\begin{align}
-\hat{H}_0^{*}(-\textbf{k}) = \hbar v k\begin{pmatrix}
\textrm{cos} \phi & e^{i \theta} \textrm{sin} \phi  \\ e^{-i \theta} \textrm{sin} \phi  & -\textrm{cos} \phi 
\end{pmatrix}.
\end{align}
For $\Delta_S=0$, the electron and hole parts of $H_{BdG}$ decouple and one can define a basis where the spinors of the electron quasiparticle are $\psi_{e1,2}=\frac{1}{\sqrt{2}}\left(\textrm{cos}\frac{\phi}{2},e^{i \theta}\textrm{sin}\frac{\phi}{2}, \pm \textrm{sin}\frac{\phi}{2}, \mp e^{i \theta}\textrm{cos}\frac{\phi}{2} ,0,0,0,0\right)^T$, in analogy with $\psi_D$, with $E=-\mu+\hbar vk_e$, and the spinors of the hole quasiparticles, $\psi_{h1,2}=\frac{1}{\sqrt{2}}\left(0,0,0,0, \textrm{sin}\frac{\phi}{2}, -e^{-i \theta}\textrm{cos}\frac{\phi}{2}, \mp \textrm{cos}\frac{\phi}{2}, \mp e^{-i \theta}\textrm{sin}\frac{\phi}{2} \right)^T$ with $E=\mu-\hbar v k_h$. We assume that $\mu \gg \Delta_S$, so that we can make the usual Andreev approximation, in which $k_e\approx k_h$, and we take the same $\theta$ and $\phi$ in the electron and hole branches. 

\begin{figure}
\includegraphics[clip=true,width=12cm]{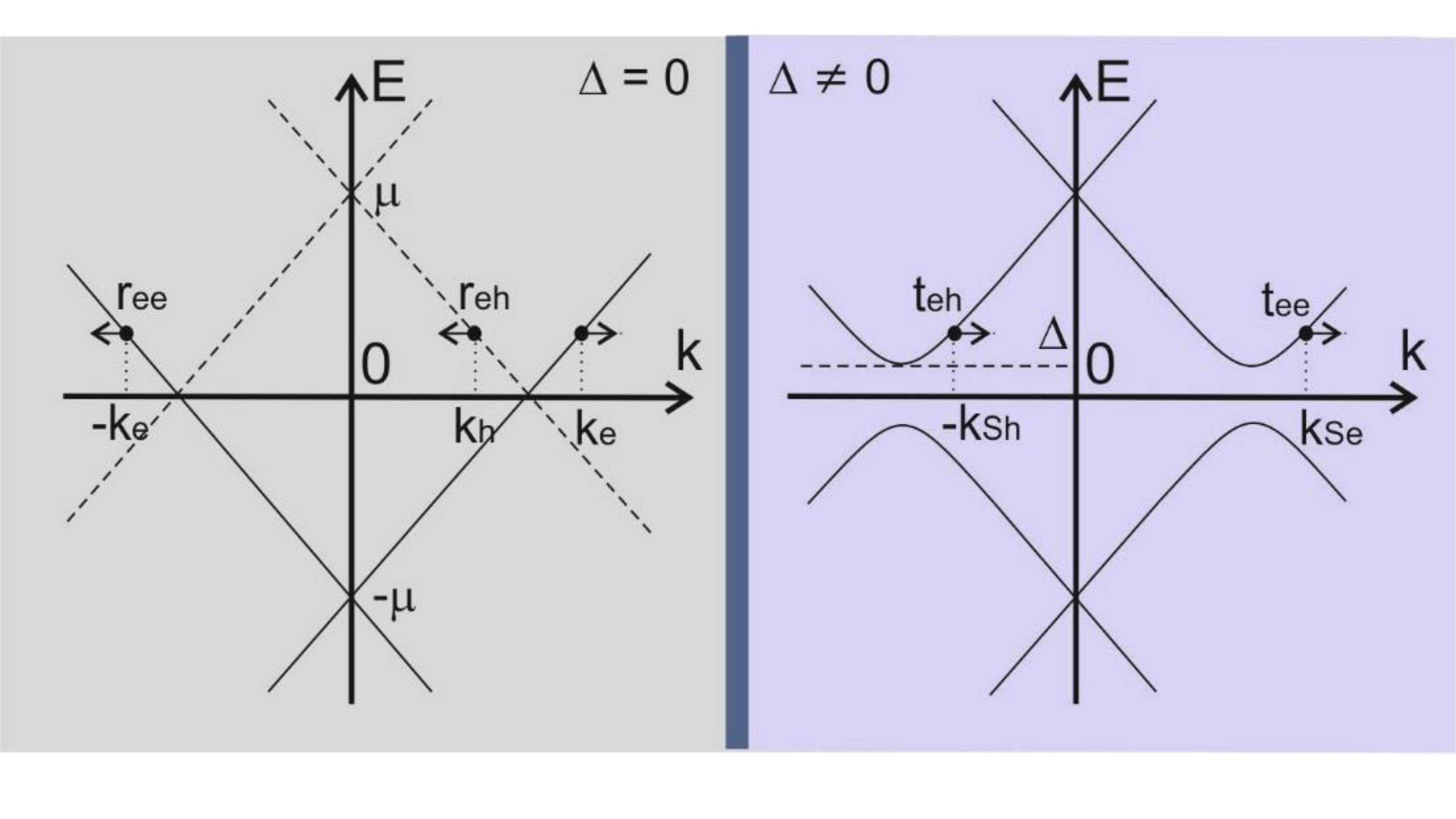}
\caption{Left: Linear dispersion of a Dirac semimetal, where the solid line represents the electron description of a Dirac cone, whereas the dashed line is the hole representation. Right: the dispersion relation of the electronlike and holelike Bogoliubov quasiparticles in a superconducting Dirac semimetal. Note, that all drawn dispersion relations are doubly degenerate since two Dirac cones exist with opposite chirality. Also indicated are the reflection and transmission processes for an electron in a Dirac semimetal (no pairing, so $\Delta=0$) travelling to the right towards an interface with the superconducting Dirac semimetal ($\Delta \neq 0$).  }
\label{Fig:Theory}
\end{figure}

$H_{BdG}$ describes the induced superconductivity inside a Dirac semimetal, such as below the Nb electrodes of the devices under study. For example, for $0<E<\mu$, the dispersion is given by $E=\sqrt{(\mu-\hbar v k)^2+\Delta^2}$, with corresponding spinors $\psi_{S}^{1,2}=\frac{\chi}{2\sqrt{E}}\left(e^{i\varphi}\textrm{cos} \frac{\phi}{2}, e^{i\varphi}e^{i\theta}\textrm{sin} \frac{\phi}{2} , \pm e^{i\varphi}\textrm{sin} \frac{\phi}{2}, \mp e^{i\varphi}e^{i\theta}\textrm{cos} \frac{\phi}{2}, -\frac{\Delta}{\chi^2} e^{i\theta}\textrm{sin}\frac{\phi}{2}, \frac{\Delta}{\chi^2}\textrm{cos}\frac{\phi}{2}, \pm \frac{\Delta}{\chi^2} e^{i\theta}\textrm{cos} \frac{\phi}{2}, \pm \frac{\Delta}{\chi^2}\textrm{sin} \frac{\phi}{2}\right)^T$ with $\chi=\sqrt{E-\mu+\hbar v k}$. At energy $E$, two electronlike and two holelike Bogoliubov quasiparticles exist for each of the two superconducting condensates, see Fig. \ref{Fig:Theory}. The corresponding wave vectors are given by $\hbar v k_{Se,h} = \mu \pm \sqrt{E^2-\Delta^2}$, so that the coefficient $\chi$ is different for the electronlike and holelike quasiparticles, $\chi_{e,h}=\sqrt{E\pm\sqrt{E^2-\Delta^2}}$. This also provides different corresponding wave functions, $\psi_{Se,h}^{1,2}$. Note, that the coefficients $\chi_{e,h}$ have an imaginary component for the energies of interest here ($E<\Delta$). This gives rise to complex wave numbers, $k_{Se,h}$, where the real parts represent plane waves and the imaginary part evanescent waves.

At the interface between a Dirac semimetal and a (proximity induced) superconducting Dirac semimetal, the reflection, transmission and Andreev reflection coefficients can be obtained by the continuity condition of the wave functions at the two sides of the interface, following the same formalism as developed for the normal metal - superconductor interface \cite{BTK}. For example, an electron in cone 1 approaching the interface on its right at an angle $\theta$ can reflect as an electron in cone 1 with coeffcient $r_{ee}$, as an electron in cone 2 ($r_{ee}'$), as a hole in cone 1 ($r_{eh}$), and as a hole in cone 2 ($r_{eh}'$). Into the superconductor the electron can transmit as an electronlike quasiparticle in the condensate of cone 1 ($t_{ee}$) and cone 2 ($t_{ee}'$), and as a holelike quasiparticle into cone 1 ($t_{eh}$) and cone 2 ($t_{eh}'$). We assume translational invariance along the $y$ and $z$ coordinates, which dictates the conservation of parallel momentum, $k_y$ and $k_z$ (see Fig. 1B of the main paper for the coordinate geometry). For electron reflection, only $k_x$ changes sign, which can be captured by taking $\theta \rightarrow \pi-\theta$. Upon Andreev reflection, all $k$-components are conserved. If we allow the chemical potential in the superconductor, $\mu_S$, to differ from the potential in the Dirac semimetal, then the angle of the transmitted particles is given by $\theta_S = \textrm{arcsin} \left(\frac{\mu}{\mu_S} \frac{\textrm{sin} \phi}{\textrm{sin} \phi_S}\textrm{sin} \theta \right)$, where $\phi_S=\textrm{arccos}\left(\frac{\mu}{\mu_S} \textrm{cos} \phi \right)$ because of the assumed conservation of momentum parallel to the interface. Transmission can occur as an electronlike and as a holelike quasiparticle. For the holelike quasiparticle $k_x$ is reversed, see Fig. \ref{Fig:Theory}, which is provided by the angle $\pi-\theta_S$. The continuity equation then reads 
\begin{eqnarray}
\psi_{e1}(\theta)+r_{ee} \psi_{e1} (\pi-\theta) + r_{ee}' \psi_{e2} (\pi-\theta) + r_{eh} \psi_{h1}(\theta)+r_{eh}' \psi_{h2} (\theta) \nonumber \\ =t_{ee} \psi_{Se1}(\theta_S) + t_{ee}' \psi_{Se2}(\theta_S) + t_{eh} \psi_{Sh1}(\pi-\theta_S) + t_{eh}' \psi_{Sh2}(\pi-\theta_S),
\label{continuity}
\end{eqnarray}
which provides 8 equations with 8 unknowns. In a similar way, the coefficients for an incoming hole in one of the cones can be calculated. For example, for cone 1,
\begin{eqnarray}
\psi_{h1}(\pi-\theta)+r_{hh} \psi_{h1} (\theta) + r_{hh}' \psi_{h2} (\theta) + r_{he} \psi_{e1}(\pi-\theta)+r_{he}' \psi_{e2} (\pi-\theta) \nonumber \\=t_{he} \psi_{Se1}(\theta_S) + t_{he}' \psi_{Se2}(\theta_S) + t_{hh} \psi_{Sh1}(\pi-\theta_S) + t_{hh}' \psi_{Sh2}(\pi-\theta_S).\label{continuity2}
\end{eqnarray}
This procedure can be repeated for particles directed towards a second interface with another superconductor (different phase factor $\varphi$). 

In the case of transport in the lateral plane of the junction ($k_z=0$) and for an angle of incidence of $\theta=0$, most of the coefficients are found to be zero. In fact, we then obtain $r_{ee} =r_{ee}'= r_{eh}'=t_{eh} =t_{eh}' =0$. In this case, the vanishing reflection coefficient for scattering from one cone to the other, $r_{ee}'=0$ is physically explained by the orthogonality of the wave functions. Within one cone the reflection coefficient, $r_{ee}$, is 0 due to the ortogonality of the wave functions for $\theta=0$ and $\theta=\pi$. The remaining Andreev coefficients are derived to be $r_{eh}^l=\frac{\Delta}{E+\sqrt{E^2-\Delta^2}} e^{-i\varphi_l}$ and $r_{he}^r= \frac{E-\sqrt{E^2-\Delta^2}}{\Delta} e^{i\varphi_r}$, where the indices $l$ and $r$ refer to the left and right interfaces respectively.

Once all the (Andreev) reflection and transmission coefficients are known, the Andreev bound state can be calculated. Here, we generalize the procedure by Kulik \cite{Kulik} and write the wave function in the Dirac semimetal between two superconducting Dirac semimetals as 
\begin{equation}
\psi=a_1 \psi_{e1}^+ + a_2 \psi_{e2}^+ + b_1 \psi_{h1}^+ + b_2 \psi_{h2}^+ +c_1 \psi_{e1}^- + c_2 \psi_{e2}^- +d_1 \psi_{h1}^- + d_2 \psi_{h2}^- \label{psi},
\end{equation}
where +(-) indices refer to right (left) propagating waves, and where the coefficients $a_1$ to $d_2$ are given by 8 continuity equations, such as $c_1 e^{-i k_e L}=r_{ee}^r a_1 e^{i k_e L}+ r_{he}^r d_1 e^{-i k_h L} + r_{ee}'^r a_2 e^{i k_e L}+ r_{he}'^r d_2 e^{-i k_h L}$. Here, $L=\frac{l}{2} \textrm{cos} \theta \textrm{sin} \phi$, where the interfaces are assumed to be at $x=\pm \frac{l}{2}$. The electron and hole wave vectors are given by $\hbar v k_{e,h}= \sqrt{\mu \mp E}$. In the end, solving the 8 equations provides the energy of the bound state as a function of the phase difference between the superconductors. If we now look at the case where $k_z=0$ and $\theta=0$ again, the bound state condition becomes $r_{eh}^l r_{he}^r = 1$ in the limit of $k_{e,h}L \ll 1$, which then gives $E_{\pm}=\pm \Delta \textrm{cos} \left(\frac{\varphi_l - \varphi_r}{2}\right)$. This Andreev bound state has a 4$\pi$ periodicity. An example of numerically calculated Andreev bound states for modes with a finite momentum parallel to the interface (nonzero $k_z$ and/or $k_y$) is given in Fig. 1 of the main text. The derivation of the bound states can easily be generalized to junctions with arbitrary electrode spacing, $L$, by keeping the separate expressions for $k_e$ and $k_h$. Examples will be given in the section below. 

The $E=0$ state around $\varphi=\pi$ of the 4$\pi$-periodic current-phase relation can be described as a Majorana state. When solving for the coefficients $a_1$ to $d_2$ in Eq. (\ref{psi}), for $k_z=k_y=0$, the coefficients can be chosen such that $\psi$ is written as a superposition of the wave functions belonging to $E_+$ and $E_-$, namely $\psi=a_1 \left(\psi_{e1}^+ + r_{he}^l \psi_{h1}^+ \right)+c_1 \left(\psi_{e1}^- + r_{he}^r \psi_{h1}^- \right)+a_2 \left(\psi_{e2}^+ + r_{he}^l \psi_{h2}^+ \right)+c_2\left(\psi_{e2}^- + r_{he}^r \psi_{h2}^- \right)$. Note, that for $\varphi_r=-\varphi_l=-\frac{\pi}{2}$, the reflection coefficients at $E=0$ simplify to $r_{he}^r=-1$ and $r_{he}^l=1$. Generalizing the procedure by Fu and Kane \cite{FuKane2008} to the spinors of the Dirac semimetal bound states, we find $\psi=\xi_1 \pm \xi_2$, where $\xi_1=\frac{1}{2}\left(1,0,0,0,0,-1,0,0 \right)^T$ corresponds to a basis vector where the coefficients obey $a_1=c_1=a_2=c_2$, and $\xi_2=\frac{1}{2}\left(0,1,0,0,1,0,0,0 \right)^T$ corresponds to $a_1=a_2=-c_1=-c_2$. By assuming continuity across the interfaces, the same spinor can be written at the superconducting side of the interface. The spatial part of the wave function at $E=0$ is given by the wave number $\hbar v k_{Se,h}=\mu \pm i \Delta$ providing an exponentially decaying factor, $e^{-x/\xi}$, where the length scale of the Majorana states $\xi = \frac{\hbar v}{\pi \Delta}$. 

\subsection{Resonances and estimated 4$\pi$ contribution}
In (semi)-metallic heterostructures, such as the Josephson junctions under study, the Fermi wavelength ($\lambda_F$) will likely be shorter than the interlayer length ($L$). For ballistic transport this can give rise to Fabry-Perot type resonances in the interlayer region. Just like in an optical system, the normal state transmission through the structure is maximal when $k_x L=n \pi$, where $k_x=k_F \textrm{cos}(\theta)$ is the momentum in the direction perpendicular to the interfaces. In the case of superconducting electrodes, the supercurrent is carried by Andreev bound states. The Andreev bound states can fully contribute to the supercurrent whenever the Andreev bound state energy lies within a normal state resonance. The electrons and holes in the Andreev bound state can stay coherent, whereas outside the normal state resonance they quickly dephase. Coherence effects from overlapping normal state resonances and Andreev bound states were, for example, shown for ballistic double-barrier Josephson junctions with a metallic interlayer \cite{Brinkman}. In that case, a normal state resonance is narrow in energy, its width given by $\gamma \hbar v_F /L$, where $\gamma$ is the angle averaged single barrier transparency. As a consequence, the transport is carried only by modes at angles very close to resonant conditions.

\begin{figure}
\includegraphics[clip=true,width=15cm]{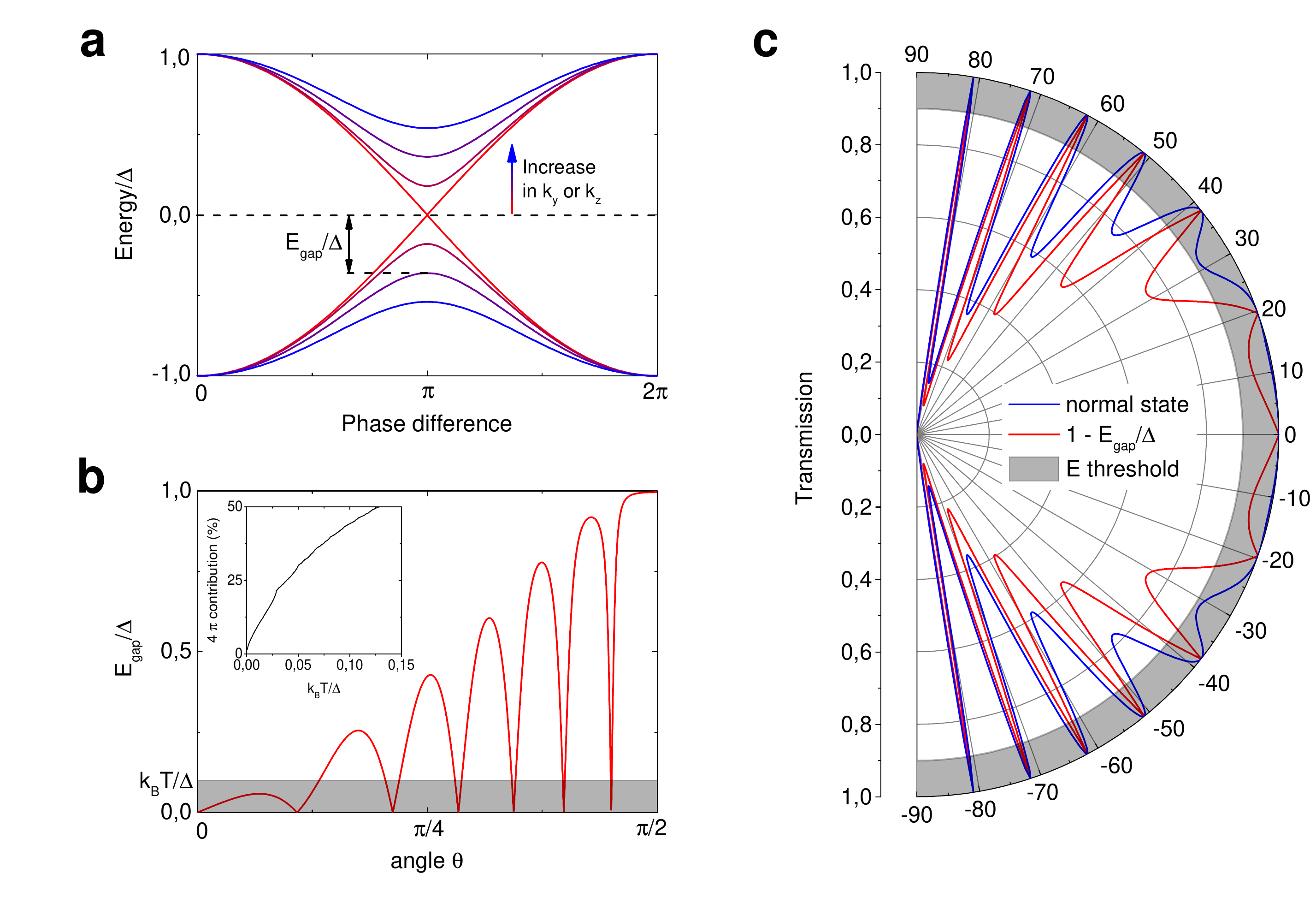}
\caption{\textbf{Transmission resonances and Andreev bound states. a,} Calculated energy of the Andreev bound states as a function of the phase difference across a superconductor - Dirac semimetal - superconductor Josephson junction. For Andreev bound states in the direction perpendicular to the interfaces, the bound states cross $E=0$ at a phase difference of $\pi$, resulting in a $4\pi$-periodicity. For nonzero values of the parallel momentum, a gap ($E_{\textrm{gap}}$) is opened in the bound state spectrum, giving a $2\pi$-periodicity. \textbf{b,} The energy gap between Andreev bound states at a phase difference of $\pi$ as a function of the angle ($\theta$) between the propagation direction and the normal to the interface. At angles for which $E_{\textrm{gap}}$ is smaller than other energy scales, such as temperature ($k_BT$), the Andreev bound states cannot be distinguished experimentally from $4\pi$-periodic bound states (gray area). For this example $k_FL=10$ and $\mu_{\textrm{el}}=2\mu_{\textrm{int}}$. Inset: percentage of angles for which $E_{\textrm{gap}}<k_B T$ as a function of the cut-off energy ($k_BT$). Here, $k_FL=157$ and $\mu_{\textrm{el}}=2\mu_{\textrm{int}}$ have been determined from the experimental parameters of the junctions. \textbf{c,} Normal state transmission resonances (blue line) occur at angles for which $k_FL \textrm{cos}(\theta)=n \pi$. These angles coincide with the angles for which the Andreev bound states cross zero energy, i.e. $E_{\textrm{gap}}=0$, or $1-E_{\textrm{gap}}/\Delta=1$ (red line). Here, $k_FL=10$ and $\mu_{\textrm{el}}=2\mu_{\textrm{int}}$.}
\label{Fig:Fig_resonance}
\end{figure}

In the present case of rather transparent interfaces (no intentional tunnel barriers), one can expect the normal state resonances to be much broader, both in energy and in angle. Because of the topological protection against back-scattering due to the spin-momentum locking in topological insulators, Dirac semimetals and Weyl semimetals, the resonances become even broader. Additionally, a very broad transmission resonance appears perpendicular to the interfaces. These effects can be obtained from the formalism of the previous section by switching off the superconductivity in the electrodes (taking $\Delta=0$) and calculating the total transmission coefficient. We then reproduce the results that have been obtained for graphene \cite{Katsnelson}, where the normal state transmission coefficient ($D=1-|r|^2$) through a graphene trilayer was obtained from
\begin{equation}
r=2i e^{i \phi} \textrm{sin} (k_x L) \frac{\textrm{sin} (\phi) -\textrm{sin} (\theta)}{e^{-ik_xL} \textrm{cos}(\phi+\theta)+e^{ik_xL}\textrm{cos}(\phi-\theta)-2i \textrm{sin}(k_xL)}.
\end{equation}
Here, $\phi=\textrm{sin}^{-1}\left[\frac{\mu_{\textrm{int}}}{\mu_{\textrm{el}}}\textrm{sin}(\theta) \right]$ is determined by the ratio of the Fermi energies in the interlayer ($\mu_{\textrm{int}}$) and electrode ($\mu_{\textrm{el}}$).
This transmission is the same for all interlayer materials with (pseudo-) spin-momentum locking, and has been plotted in Fig. \ref{Fig:Fig_resonance}c for a chemical potential difference $\mu_{\textrm{el}}=2\mu_{\textrm{int}}$. In addition to the broad transmission resonance at $\theta=0$, normal state resonances appear for every $k_F L \textrm{cos}(\theta)=n \pi$. The transmission peak around the perpendicular mode is broad because backscattering is suppressed with a factor that scales with the cosine of the angle.

When we now consider superconducting electrodes, the Andreev bound state spectrum is calculated as outlined in the previous section. It was described that the perpendicular Andreev bound state is 4$\pi$ periodic, and this still holds also for longer junctions with multiple normal state resonances. Moving away from the perpendicular direction, a gap is opened in the Andreev bound state spectrum ($E_{\textrm{gap}}$) as an avoided level crossing, as shown in Fig. \ref{Fig:Fig_resonance}a. However, this gap closes again for every angle that has a normal state resonance (as was noticed before in the context of 3D topological insulators \cite{Snelder}), making these Andreev bound states 4$\pi$ periodic again. An example of $E_{\textrm{gap}}$ at a phase difference of $\pi$ is given in Fig. \ref{Fig:Fig_resonance}b. In Fig. \ref{Fig:Fig_resonance}c it can be seen that the 4$\pi$ Andreev bound states (for which $1-E_{\textrm{gap}}/\Delta = 1$) occur at the same angles as the normal state resonances.

In order to estimate the ratio of 4$\pi$- and 2$\pi$-periodic bound states in the spectrum one should realize that 2$\pi$-periodic Andreev bound states with very small gaps cannot be distinguished experimentally from 4$\pi$-periodic Andreev bound states if $E_{\textrm{gap}}$ is smaller than other relevant energy scales such as $k_BT$ (giving thermal noise broadening of the levels) or $eV_{\textrm{bias}}$ (providing a probability for Landau-Zener tunneling between Andreev levels across the gap). In practice, to estimate the observable fraction of the 4$\pi$ contribution to the supercurrent, we can define an energy cut-off, e.g. $k_BT$, as shown by the gray area in Fig. \ref{Fig:Fig_resonance}b, and count the fraction of Andreev bound states with $E_{\textrm{gap}}<k_BT$. Because of the broad normal state transmission resonances of topological systems (especially for the forward direction), this fraction can be quite large. It is not always possible to increase temperature to enhance the relative $4\pi$ contribution because the critical current strongly decays with temperature, which limits the observability of the $4\pi$ contribution in itself.

For the Dirac semimetal Josephson junctions under discussion in the main text we can make an estimation of the number of transmission resonances by taking $E_F=16$ meV in the interlayer. Since $E^2=(\hbar v_x k_x)^2+(\hbar v_y k_y)^2+(\hbar v_z k_z)^2$, we obtain a maximal $k_x^{\textrm{max}}=\frac{E_F}{\hbar v_x}=3.2 \times 10^8$ m$^{-1}$ for $v_x=8 \times 10^4$ m/s. The number of resonances, $\frac{k_x^{\textrm{max}}L}{\pi}$ is then about 50 for a junction length of 500 nm. Note, that we treat the transverse momenta $k_y$ and $k_z$ as continuous, whereas these are quantized in practice due to quantum confinement. However, the number of modes in the $y$-direction can be estimated for a width of 3 $\mu$m and $v_y=10^6$ m/s as $\frac{k_y^{\textrm{max}}W}{\pi}=25$, which is still reasonably large. For these parameters, the observable fraction of $4\pi$ Andreev bound states is depicted in the inset of Fig.  \ref{Fig:Fig_resonance}b as a function of the measurement temperature. In order to match the experimentally observed fraction of about 20\%, $k_BT/\Delta$ needs to be about 0.03. For a proximity induced $\Delta$ of about 0.5 meV, this would correspond to a temperature of about 0.17 K. This seems to be somewhat larger than the actual measurement temperature, but one should take into account that this is only a qualitative estimate with quite some uncertainty in the values of the parameters. For example, for a ratio of the chemical potentials closer to unity, 20\% of $4\pi$ contribution would be reached at lower temperatures already.

\FloatBarrier
\subsection{Field dependence of the critical current}

\begin{figure}
\includegraphics[clip=true,width=10cm]{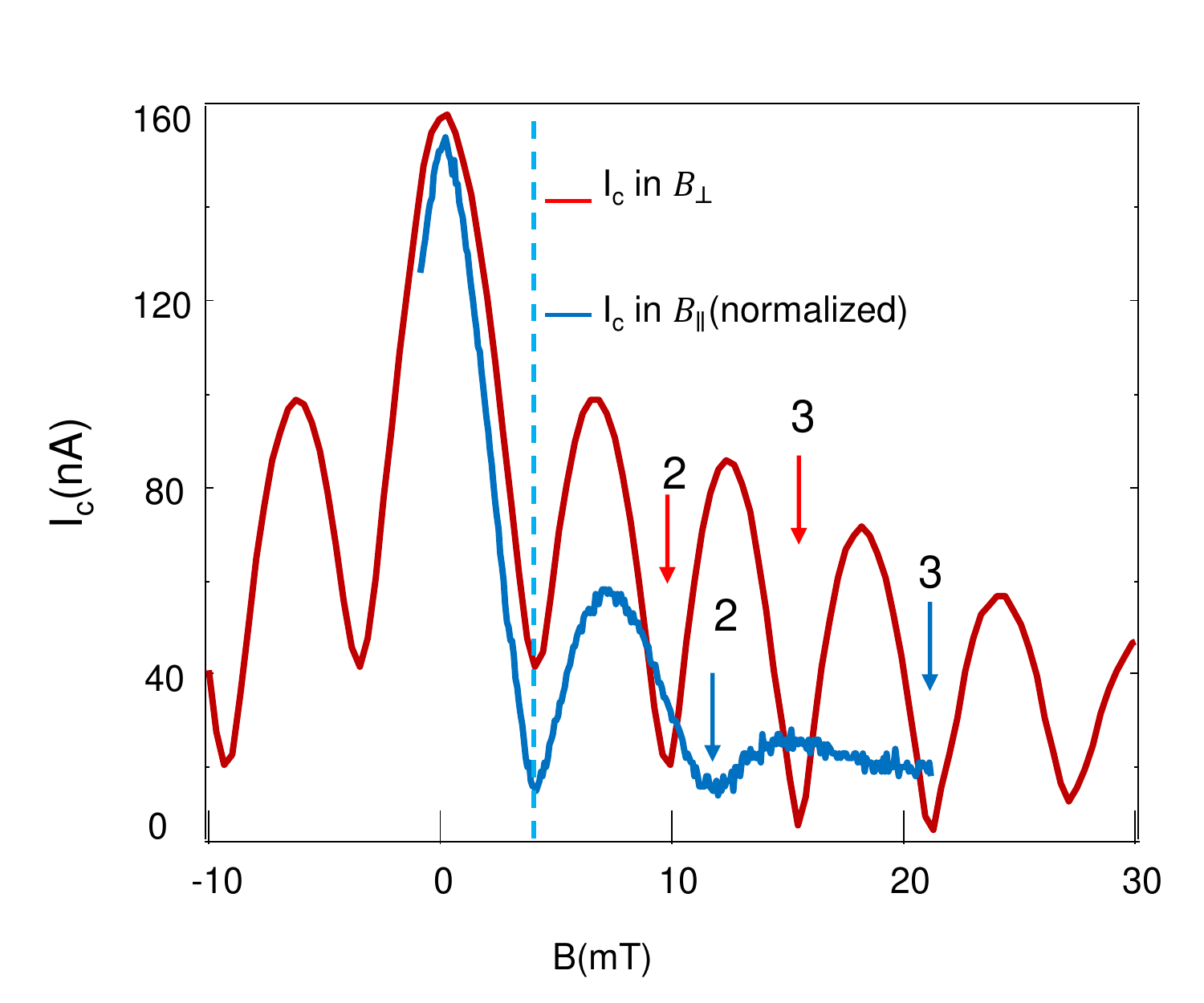}
\caption{\textbf{Critical current modulation by magnetic field} For an out-of-plane magnetic field \cite{Novoselov}, the critical current of the Josephson junctions are modulated in a Fraunhofer-type fashion. For an in-plane magnetic field (blue trace) the critical current modulation is qualitatively different, ruling out an out-of-plane field component as the cause of the modulation for parallel field. The dependencies are shown for a 800 nm long junction, the same junction for which the parallel field dependence is shown in the main paper in Fig. 4. In order to compare the dependencies (e.g. the positions of the minima), the magnetic field scale of the parallel orientation has been normalized to let the first minimum coincide, indicated by the blue dashed line.}
\label{Fig:Fig_Fraunhofer}
\end{figure}

In Fig. \ref{Fig:Fig_Fraunhofer} we show the magnetic field dependence of the critical current. For the out-of-plane field direction, the critical current modulation resembles the so-called Fraunhofer pattern, in which the critical current amplitude oscillates as a sinc function with perpendicular field strength. The measured period of $I_c$ of $\Delta B=857$ $\mu$T corresponds to an area $A = \Phi_0/\Delta B = 2.33$ $\mu$m$^2$. This area is about 1.5 times larger than the actual area of the junction, which can be attributed to the penetration depth of the Nb electrodes. For a magnetic field applied parallel to the junction (and parallel to the current direction), the modulation of the critical field is qualitatively different. This can be seen in Fig. \ref{Fig:Fig_Fraunhofer}, where the magnetic field scale of the parallel applied field is normalized so that the first minimum in the critical current coincides with that for the perpendicular field. The qualitative difference shows that the in-plane magnetic field dependence cannot be explained as a Fraunhofer dependence coming from a small out-of-plane component of the magnetic field. Rather, as is shown by the fit in the main text of the paper, the in-plane field dependence of the critical current can be explained by a shift of the bulk electron Fermi surface, leading to a finite momentum of the Cooper pairs. The scale of the decay and the oscillation period in magnetic field in this case are related \cite{Born} and can be used to determine the elastic mean free path for the 800 nm junctions that are on the border between ballistic and diffusive. When looking at Fig. 5 of Ref. \onlinecite{Born}, it can be seen that both the oscillation period and the decay length are proportional to $\xi_H= \frac{\hbar v_F}{g\mu_B B}$ (as we also observe in our measurement) when the mean free path and the coherence length are of the same order of magnitude.